\documentclass[useAMS,usenatbib]{mn2e}

\usepackage{times}
\usepackage{amsmath}
\usepackage{amssymb}
\usepackage{graphicx}
\usepackage{xspace}
\usepackage{subfigure}

\newcommand{\abs}[1]{\left|{#1}\right|}
\newcommand{\chem}[1]{\ensuremath{\mathrm{#1}}}
\newcommand{\pdiff}[2][{}]{\ensuremath{\partial_{#2}\ifx\\#1\else^{#1}\fi}}
\newcommand{\pdsec}[2]{\ensuremath{\frac{\partial^2{#1}}{\partial{#2}^2}}}
\newcommand{\dsec}[2]{\ensuremath{\frac{d^2{#1}}{d{#2}^2}}}
\newcommand{\deriv}[2]{\ensuremath{\frac{d{#1}}{d{#2}}}}
\newcommand{\pderiv}[2]{\ensuremath{\frac{\partial{#1}}{\partial{#2}}}}
\newcommand{\jacpoly}[2]{\ensuremath{\mathop{P^{1{,}1}_{#1}\!}\left(#2\right)}}
\renewcommand{\vec}[1]{\ensuremath{\mbox{\boldmath${#1}$\unboldmath}}}
\newcommand{\unit}[2][{}]{\ensuremath{\ifx\\#1\else#1\,\fi\mathrm{#2}}}

\newcommand{\alphadip}{\alpha_\mathrm{dip}}
\newcommand{\Bdip}{B_\mathrm{dip}}
\newcommand{\Bt}{B_\theta}
\newcommand{\Br}{B_r}
\newcommand{\BZ}{B_Z}
\newcommand{\Brho}{B_\rho}
\newcommand{\Beq}{B_0}
\newcommand{\Bteq}{B_{\theta0}}

\newcommand{\kb}{k_b}
\newcommand{\lsc}{\ell}
\newcommand{\RP}{\mathrm{R_P}}
\newcommand{\RJ}{\mathrm{R_J}}
\newcommand{\RS}{\mathrm{R_S}}
\newcommand{\RMP}{\mathrm{R_{MP}}}
\newcommand{\RT}{\mathrm{R_T}}

\newcommand{\Cassini}{\textit{Cassini}\xspace}
\newcommand{\Voyager}{\textit{Voyager}\xspace}
\newcommand{\Galileo}{\textit{Galileo}\xspace}
\newcommand{\CAPS}{\renewcommand{\CAPS}{CAPS\xspace}\Cassini Plasma Spectrometer (CAPS)\xspace}
\newcommand{\MAG}{\renewcommand{\MAG}{MAG\xspace}\Cassini magnetometer (MAG)\xspace}
\newcommand{\MIMI}{\renewcommand{\MIMI}{MIMI\xspace}\Cassini Magnetospheric Imaging Instrument (MIMI)\xspace}
\newcommand{\INCA}{\renewcommand{\INCA}{INCA\xspace}ion-neutral camera (INCA)\xspace}
\newcommand{\PLS}{\renewcommand{\PLS}{PLS\xspace}\Voyager plasma science (PLS)\xspace}
\newcommand{\SLT}{\renewcommand{\SLT}{SLT\xspace}Saturn local time (SLT)\xspace}

\newcommand{\rev}[1]{\renewcommand{\rev}[1]{Rev~##1\xspace}Revolution~#1 (Rev~#1)\xspace}

\newcommand{\Eq}[1]{Eq.~({#1})}
\newcommand{\Eqs}[1]{Eqs.~({#1})}
\newcommand{\Fig}[1]{Fig.~{#1}}
\newcommand{\Table}[1]{Table~{#1}}

\newcommand{\comment}[1]{}

\newlength{\figwidth}
\figwidth=\columnwidth 

\title[A model of force balance  
in Saturn's magnetodisc]{A model of force balance  
in Saturn's magnetodisc}
\author[N. Achilleos, P. Guio and C. S. Arridge]{N. Achilleos$^{1,2}$\thanks{E-mail:
nick@apl.ucl.ac.uk}, P. Guio$^{1,2}$ and C. S. Arridge$^{3,2}$ \\
$^{1}$Department of Physics and Astronomy, University College London, Gower St., London, WC1E 6BT, U.K. \\
$^{2}$The Centre for Planetary Sciences at UCL/Birkbeck, Gower St., London, WC1E 6BT, U.K. \\
$^{3}$Mullard Space Science Laboratory, Department of Space and Climate Physics, UCL, Holmbury St. Mary, Dorking, Surrey, RH5 6NT, U.K.}

\begin{document}

\date{Accepted 2009 October 9. Received 2009 October 7; in original form 2009 September 8}

\pagerange{\pageref{firstpage}--\pageref{lastpage}} \pubyear{2009}

\maketitle

\label{firstpage}

\begin{abstract}
We present calculations of magnetic potential functions associated with the perturbation 
of Saturn's planetary magnetic field by a rotating, equatorially-situated disc of plasma. Such structures are 
central to the dynamics of the rapidly rotating magnetospheres of Saturn and Jupiter. They are `fed' 
internally by sources of plasma from moons such as Enceladus (Saturn) and Io (Jupiter). For these 
models, we use a scaled form of Caudal's Euler potentials for the Jovian magnetodisc field 
\citep{caudal}. In this formalism, the magnetic field is assumed to be azimuthally symmetric about the planet's axis of rotation, 
and plasma temperature is constant along a field line. We perturb the dipole potential (`homogeneous' solution) 
by using simplified distributions of plasma pressure and angular velocity for both planets, based on 
observations by the \Cassini (Saturn)  and \Voyager (Jupiter) spacecraft. Our results quantify the degree of 
radial `stretching' exerted on the dipolar field lines through the plasma's rotational motion and pressure. 
A simplified version of the field model, the `homogeneous disc',  can be used to easily estimate
the distance of transition in the outer magnetosphere between pressure-dominated and 
centrifugally-dominated disc structure.
We comment on the degree of equatorial confinement as represented by the scale height associated with 
disc ions of varying mass and temperature.  For the case of Saturn, we identify the principal
forces which contribute to the magnetodisc current and make comparisons between the field structure
predicted by the model and magnetic field measurements from the \Cassini spacecraft. 
For the case of Jupiter, we reproduce Caudal's original calculation 
in order to validate our model implementation. We also show that compared to Saturn, where plasma
pressure gradient is, on average, weaker than centrifugal force, the outer plasmadisc of Jupiter is
clearly a pressure-dominated structure.
\end{abstract} 
\begin{keywords}
(magnetohydrodynamics) MHD --- plasmas --- methods: numerical --- planets and satellites: general  
\end{keywords}

\section{Introduction}
\label{sec:intro}
Jupiter and Saturn are not only the largest planets in our Solar system, they are also the most rapid rotators.
\cite{gledhill1967} first pointed out the important consequences of these properties for Jupiter's magnetosphere.
The rotational period of the planet is approximately \unit[10]{h}, and as a result the gravitational ($F_g$) and centrifugal ($F_c$) 
forces associated with corotating plasma in Jupiter's magnetosphere are equal at an equatorial distance of
 $\sim\unit[2.3]{\RJ}$
from the planet's centre (here, we denote Jupiter's radius as $\RJ\approx\unit[71000]{km}$). At the orbit
of Io, situated at \unit[6]{\RJ}, centrifugal force exceeds gravitational by a factor of nearly 20. Saturn's radius 
($\RS\approx\unit[60000]{km}$) and rotational
period ($\sim\unit[10.75]{h}$) modify these distances to \unit[1.7]{\RS} ($F_g=F_c$)
and  \unit[4.7]{\RS} ($F_g\approx F_c/20$), the latter being \unit[0.7]{\RS}
outside the orbital radius of the icy moon Enceladus. Evidently, centrifugal force is an important factor
for determining the structure of the outer magnetospheres of these planets.

\cite{gledhill1967} showed that the action of centrifugal force in Jupiter's
rapidly-rotating magnetosphere tends to confine magnetospheric plasma towards the equatorial plane, 
where the planet's assumed dipolar field lines reach their maximum radial distances. A disc-like magnetospheric
structure was thus anticipated, and indeed observed by the first spacecraft to visit the Jovian system,
Pioneers~10 and 11 \citep{smith1974,smith1975}. The near-equatorial magnetic field structure seen by these spacecraft
was very different from that of a rotating dipole throughout the region referred to as the middle magnetosphere, situated
at distances $\sim\unit[20\mbox{--}50]{\RJ}$. The magnetometer observations in this region showed a periodic pattern of 
largely radial field direction alternating with intervals having a north-south (meridionally-directed) field. These data
were interpreted as periodic encounters with a rotating, disc-like current sheet. The highly radial field in this picture
is a signature of magnetic field lines resembling a dipole pattern that has been radially `stretched' outwards near the
magnetic equatorial plane. Such a magnetic geometry would be associated with an inward Lorentz force 
$\sim\vec{J_{\phi}}\times\vec{B}$, part of which is required to provide the centripetal acceleration for the 
rotating magnetospheric plasma (here \vec{J_{\phi}} denotes
azimuthal current density and \vec{B} magnetic field).

Observations by \Galileo \citep{kivelson1997} confirmed the persistence of this magnetodisc structure, and examined
its response to changing solar wind conditions. During the late inbound pass of the \Galileo insertion orbit, the magnetic
field measurements indicated that a strong compression of the magnetosphere had taken place \citep{kivelson1997}.
This compression witnessed by \Galileo resulted in an increase of the meridional field $\Bt$ by a factor of $\sim2$ when compared with the
data from the Pioneer~10 outbound segment over the middle magnetospheric
region at \unit[30\mbox{--}50]{\RJ}. Like \Galileo, 
Pioneer~10 outbound was a near-equatorial swathe situated at a local time near dawn. Unlike \Galileo,
however, the Pioneer~10 observations, acquired 22~years earlier, indicated a relatively quiescent magnetosphere. 
The conclusion was that the magnetospheric compression at the time of the
\Galileo insertion had caused an increase in meridional field $\Bt$ 
by squeezing the magnetic flux threading the magnetospheric plasma into a smaller volume (the change in location of the dawn
magnetopause was inferred to be \unit[40]{\RJ} inward). In addition, the periodic field signatures in 
meridional and radial field ($\Bt$ and $\Br$)
seen by \Galileo indicated a thicker plasma sheet within the 
magnetodisc structure, as one would expect for a strongly-compressed magnetosphere.

The analogous behaviour for the magnetodisc at Saturn was explored by \citet{arridge2008}, who took advantage of many orbits of magnetometer
measurements from the \Cassini spacecraft in order to investigate the relationship between magnetosphere size (as represented by the subsolar 
magnetopause standoff distance $\RMP$)
and the degree to which the radial field $\Br$ dominated the field measurements seen in the outer magnetosphere.
For the observations considered in this study, \Cassini was typically situated
on near-equatorial orbits just outside, and south of, the current sheet. This analysis revealed that, under conditions of low 
solar wind dynamic pressure (corresponding to
$\RMP > \unit[23]{\RS}$), the magnetic field due to 
Saturn's ring current (i.e.\ azimuthal current) dominates the planetary internal field 
in the outer magnetosphere, with the combination of
the two producing the magnetodisc structure. For a more compressed magnetosphere, however, the dayside field became strongly dipolar, with magnetodisc geometry surviving only
on the magnetosphere's nightside and flanks. The Kronian magnetodisc may thus essentially disappear on the dayside under appropriate conditions, and is therefore 
even more sensitive than Jupiter's magnetodisc in response to upstream solar wind conditions \citep{arridge2008}.
\citet{bunce2008} arrived at similar conclusions by modelling the response of the ring current and its magnetic moment
for different magnetospheric configurations. This was done by the
application of an empirical field model (CAN disc) to different orbits of \Cassini magnetometer data. We describe 
this field model in further detail below.

\citet{connerney1981,connerney1983} provided the first detailed modelling 
of the ring current which supports the magnetodisc field at Saturn, and applied this model to the magnetometer data 
from the \Voyager spacecraft encounters. The magnetic field in the Connerney, or CAN, 
model is computed by assuming, {\em a priori\/}, 
an azimuthally symmetric distribution of current which is confined to an annular disc of uniform thickness, 
extending between inner and outer edges at specified radial distances. Current density $J_{\phi}$ in this 
model is assumed to be inversely proportional to cylindrical radial distance 
($J_{\phi} \propto 1/\rho$).  This formalism
has been employed in several studies of the structure of Saturn's ring current, usually based on fitting {\em in situ\/} magnetic field measurements
from spacecraft (e.g.\ \citet{connerney1983, giampieri2004, bunce2007}). 

The study by \citet{bunce2007} emphasised that the current which flows in the magnetodisc current sheet is a macroscopic manifestation of the microscopic drift motions 
of charged particles in the plasma. These authors examined the contribution of two types of azimuthal particle drift 
to the magnetic moment of the ring current: 
(i) the magnetic gradient drift exhibited by particles of finite thermal energy whose guiding centre moves in response to changes in field strength
experienced during individual gyrations, (ii) the inertial drift associated with the centrifugal force in a frame which corotates 
with the local plasma flow.
They showed that, for typical magnetospheric conditions at Saturn, the heavier (water-group) ions may generate a much stronger inertial
current at distances beyond $\sim\unit[10]{\RS}$ due to their rotational kinetic energy exceeding typical thermal energy. 

Theoretical and empirical magnetic field models for the ring current at Jupiter and Saturn proposed by various authors
(e.g. \citet{gleeson1976,goertz1976,connerney1981,connerney1983}) have proved to be valuable tools for determining the global
length scales and intensity of the current which supports the magnetodisc field structure. \citet{caudal} pointed out that the {\em a priori\/}
current distributions used in such models cannot be used to infer, unambiguously, the dynamical properties of the plasma in which the current flows. In particular,
determining the relative importance of the plasma pressure gradient and centrifugal
forces in generating the plasma current and magnetodisc field, requires a different approach which incorporates a knowledge of the plasma properties. 

\citet{caudal} developed a formalism
in which Jupiter's magnetic field structure was modelled by solving a
magnetostatic equation representing dynamical equilibrium, i.e.\ a uniformly zero 
vector sum for all of the aforementioned forces
throughout a specified region. This solution was then used to infer the global distribution of current which was consistent with the
equatorial distribution of plasma properties such as angular velocity, temperature, density and composition. \citet{caudal} used
observations by the \Voyager spacecraft \citep{connerney1981,bagenal1981,krimigis1981}
to constrain this equatorial plasma information, which, in his formalism, acts as a boundary condition
for inferring the global plasma properties.
The resulting current distribution
from such a calculation has a more realistic global structure than the uniformly thick, annular disc used in the empirical models. By including the effects
of both plasma thermal pressure and centrifugal force in his formalism, \citet{caudal} naturally extended previous investigations of the distortion of the 
planetary magnetic field which assumed a cold plasma with negligible thermal energy compared to rotational kinetic energy (e.g. \citet{hill.carbary1978}).

The main purpose of this paper is to adapt the formalism by \citet{caudal} in order to model the magnetodisc of Saturn. 
For the required equatorial plasma properties,
we use the latest observations by \CAPS \citep{young2004} 
and \MIMI \citep{krimigis2004}. The framework,
assumptions and inputs for the model are summarised in \S\ref{sec:modelframe}. For the sake of completeness, we provide a derivation of the magnetostatic
solution cited by \citet{caudal} in Appendix~\ref{sec:mdiscsoln}. This derivation is not published elsewhere, to the best of our knowledge. Its inclusion here serves
as a starting point for discussion of a toy model for the plasmadisc 
described in \S\ref{sec:zdisc}. This model has a very simplified structure in terms of
its plasma properties, but serves as a useful illustration of the competition between plasma pressure and centrifugal forces in determining magnetodisc structure.
Detailed magnetodisc models for Saturn are presented in \S\ref{sec:saturnmodel} and compared with \MAG data from equatorial and high-latitude
orbits. A description of \MAG 
is given in \citet{dougherty2004}. 
These model outputs are also compared with those of the best-fitting CAN discs. We conclude with a summary and discussion in \S\ref{sec:summary}.

\section{Model Framework}
\label{sec:modelframe}
\subsection{Magnetic Field Geometry and Force Balance}
We adopt the formalism of \citet{caudal}, and express the magnetic field components associated with an axially 
symmetric plasma distribution as gradients of a magnetic Euler potential
$\alpha$. 
The value of $\alpha$ is constant along any magnetic field line. It is also constant over
any axisymmetric shell of field lines (flux shell). The change in $\alpha$ between 
flux shells is simply related to the magnetic flux contained between them (i.e.\ it is a flux function,
see \S\ref{sec:coldplasma}).
With this assumption, the magnetic field radial component $\Br$ and
meridional component $\Bt$ are 
\begin{align}
\Br & = \frac{1}{r^2\sin\theta}\pderiv{\alpha}{\theta}, \notag \\
\Bt & = -\frac{1}{r\sin\theta}\pderiv{\alpha}{r},  
\label{eq:field_compts}  
\end{align} 
where $\theta$ denotes colatitude with respect to the planetary rotation axis (assumed coincident with the magnetic axis), 
and $r$ is radial distance 
from planet centre (in units of planetary radii). The unit of $\alpha$ in our `normalised' system is 
$\Beq a$, the product of the equatorial magnetic field $\Beq$ at the planet surface and the planet radius $a$. 
The adopted values and corresponding scales for relevant physical quantities 
at both Jupiter and Saturn are shown in 
\Table{\ref{scales_table}} and 
\Table{\ref{scales_appendix_table}} (Appendix~\ref{scales_appendix}).
Unless otherwise stated, we shall use this dimensionless
form of Caudal's original equations in order to easily compare the degree to which different plasma discs may distort the
internal field of their parent planets (see also \citet{vasyliunas2008} and 
\S\ref{sec:avg_mdisc}).


\begin{table*}
\caption{Physical units used in the normalised (dimensionless) system for both planets.}
\label{scales_table}
\begin{tabular}[b]{rrrrrr}
\hline 
Planet & Radius ($a$) & Magnetic Field ($\Beq$) & Magnetic Flux ($\Beq a^2$)
 & Pressure ($\Beq^2/\mu_0$)  &  Angular Velocity ($\omega_0$) \\
\hline 
Saturn & \unit[60280]{km} &                   \unit[21160]{nT} &
\unit[77]{GWb}  &          \unit[0.00036]{Pa}  &
\unit[2\pi/10.78]{rad\,h^{-1}}\\
Jupiter & \unit[71492]{km} &                 \unit[428000]{nT} &
\unit[2187]{GWb}  &      
\unit[0.146]{Pa}  & \unit[2\pi/9.925]{rad\,h^{-1}}\\
\hline
\end{tabular}  
\end{table*}

\citet{caudal} examined the condition of general force balance in the rotating plasma
\begin{align}
\vec{J} \times \vec{B} = \vec{\nabla} P - n m_{i} \omega^2 \rho \vec{e}_\rho,
\label{eq:force_balance}
\end{align}
where \vec{J} is current density, \vec{B} is magnetic field and $\rho = r \sin \theta$ is cylindrical radial 
distance from the axis ($\vec{e}_\rho$ is the corresponding unit vector).
Plasma properties are pressure $P$, temperature $T$ (assumed isotropic and constant along field lines), 
ion number density $n$, mean ion mass $m_{i}$ and
angular velocity $\omega$. 
This equation represents balance 
between magnetic force on the left side, and pressure gradient plus centrifugal force on the right.
We have not included the minor contribution to plasma mass from the electrons, but do include their
contribution to plasma pressure.

\citet{caudal} 
used the definitions of field and current expressed as functions of $\alpha$.
When these expressions are substituted into the force balance condition 
(\Eq{\ref{eq:force_balance}}), the result is
the following partial differential equation
\begin{align}
\pdsec{\alpha}{r} + \frac{1-\mu^2}{r^2} \pdsec{\alpha}{\mu}  =
-g(r,\mu,\alpha),
\label{eq:caudal_de}
\end{align}
where $\mu$ is the cosine of colatitude i.e.\ $\mu = \cos\theta$.

The `source function' $g$ is determined by the global distribution of plasma pressure and angular velocity.
\citet{caudal} pointed out that $g$ could be used to derive the azimuthal current density
$J_{\phi}$ according to 
\begin{align}
J_{\phi} (r,\mu) = \frac{g(r,\mu)}{r\sin\theta} = \frac{g(r,\mu)}{\rho}.    
\label{eq:jphifromg}
\end{align}
Force balance in the direction parallel to the magnetic field implies that the \emph{global} values of these quantities are derivable from their 
equatorial values and the shape of the magnetic field lines. This is why there is a general
dependence of $g$ upon $\alpha$.
\citet{caudal} derived an analytical expression which could be used to calculate the solution for $\alpha$.
We have included a full derivation of this expression in   
Appendix~\ref{sec:mdiscsoln}.
The form involves the
use of Jacobi polynomials, which occur as solutions of the homogeneous
version of \Eq{\ref{eq:caudal_de}}. An 
important solution in this class is the dipole potential $\alphadip = (1-\mu^2)/r$. In practice, we start with a pure 
dipole potential and then `perturb' it using Caudal's iterative method: at every iteration, the 
 solution 
$\alpha_n$ is used to evaluate $g$ and thus the `next' solution $\alpha_{n+1}$. We stopped our calculations when the 
difference between successive iterations was at most \unit[0.5]{per\,cent}. We describe the various inputs used for our Saturn model
calculations in \S\ref{sec:modelinputs}. These are based on a variety of observational studies employing data taken by
the \Cassini spacecraft. Before investigating these Caudalian disc models for Saturn, we shall 
examine
a simple toy model which may be used to predict the effect of a 
rotating plasma disc upon the zeroth-order (largest-scale) 
perturbation to the dipole potential.

\subsection{Toy Model for a Planetary Magnetodisc}
\label{sec:zdisc}
We begin our investigation by examining a very simplified model of disc structure. In this model, we assume
that the disc has a uniform plasma $\beta$ parameter denoted $\beta_h$, 
associated with the thermal energy of a hot population. We also assume the presence of an isothermal cold population
containing most of the plasma mass, but a negligibly small fraction of the
total pressure, with uniform plasma $\beta$
denoted $\beta_c$. Proceeding under these assumptions, it is straightforward to show that
\citet{caudal}'s expression for the plasma source function may be written as
\begin{align}
g(r,\mu,\alpha) = \rho^{2} \deriv{P_{h0}}{\alpha} + 
\rho^2 \exp\left(\frac{\rho^2-\rho_0^2}{2\lsc^2}\right) \frac{P_{c0}}{\lsc^2 \Bteq},
\label{eq:zdisc_g}
\end{align}
where $P_{h}$ and $P_{c}$ denote hot and cold plasma pressure, and the subscript 0 is used to refer to the quantity evaluated 
at the equatorial
crossing point of the magnetic field line, i.e.\ the magnetically conjugate point for which $\mu = 0$. 
The relation between source function and current density \Eq{\ref{eq:jphifromg}} allows the identification 
of the two terms on the right-hand side of \Eq{\ref{eq:zdisc_g}} as quantities proportional to the individual
contributions to total current
density which arise from hot pressure gradient, and from centrifugal force.
Following \citet{caudal}, 
we assume that the hot component exhibits
uniform pressure  $P_{h}$ all the way along a given field line, while the cold component's pressure is concentrated towards the 
equatorial plane,
according to the exponential factor in \Eq{\ref{eq:zdisc_g}}. The symbol $\lsc$ thus represents a 
scale length associated with the cylindrical
radial distance $\rho$, and is defined by \citep{caudal}
\begin{align}
\lsc^2 = \frac{2\kb T}{m_i\omega^2a^2}        
\label{eq:ldefn}
\end{align}
for a quasi-neutral plasma containing singly charged ions and electrons. $a$ represents planetary radius and is
used here to transform to our normalised system (see \Table{\ref{scales_table}}). 

It is worth emphasising here that this expression for 
$\lsc$ arises from \citet{caudal}'s mathematical
treatment of general force balance, but it is also a natural consequence of field-aligned force
balance for both ions and electrons in a quasi-neutral plasma. Ions and electrons in this formalism
are implicitly subject to a simplified ambipolar electric field which acts to distribute both types of particle
with an equal scale length 
$\lsc$ given by \Eq{\ref{eq:ldefn}}. We can see how this arises by
considering the equations for
field-aligned force balance, for both ions and electrons, which take into account particle
pressure, centrifugal force (projected along the magnetic field direction) and the presence of the ambipolar
electric potential 
$\Phi_{\|}$. This derivation of the scale length $\lsc$ is summarised in 
Appendix~\ref{sec:scale_length}.

Using this definition of plasma scale length, it can be shown that the exponential
factor appearing in \Eq{\ref{eq:zdisc_g}} has an argument which contains the ratio of an ion's kinetic energy of rotation
to its thermal energy. Thus our hot plasma component is defined by ion thermal energies which are large compared to the
kinetic energy of rotation at angular velocity $\omega$ and a consequent scale length 
which is effectively infinite (large compared
to magnetospheric flux tube length). On the other hand, the cold plasma component contains ions with much smaller thermal energies,
which cannot compete as effectively with the centrifugal potential in maintaining plasma at locations high above the equatorial plane.
We shall see in the later sections that the typical scale lengths for the cold plasma at Saturn are a few planetary radii, a distance which is
small compared to the flux tube lengths in the planet's outer magnetosphere.

If we now make the assumption for the global magnetic 
field $\Bteq = \rho_0^{-\chi}$ (where e.g.\ $\chi=3$ for a dipolar geometry), 
then by definition
the dependence of the normalised
magnetic pressure along the equator is given by $\frac{1}{2}\rho_0^{-2\chi}$. It follows that
$P_{h0} = \frac{1}{2}\beta_h\rho_0^{-2\chi}$ and $P_{c0} = \frac{1}{2}\beta_c\rho_0^{-2\chi}$.
Replacing the operator $\frac{d}{d\alpha}$ with the equivalent 
$\frac{-1}{\Bteq\rho_0} \frac{d}{d\rho_0}$ and assuming
uniform $T$ and $\omega$, we may transform
the expression for the plasma source function from \Eq{\ref{eq:zdisc_g}} into the following form:
\begin{align}
g(r,\mu,\alpha) = \rho^2 \rho_0^{-(\chi+2)} \left[ \beta_h \chi + 
\frac{\beta_c\rho^2}{2\lsc^2} \exp\left(\frac{\rho^2-\rho_0^2}{2\lsc^2}\right) \right].       
\label{eq:zdisc_g_trans}
\end{align}
It is straightforward to show, using \Eq{\ref{eq:ldefn}}, that the term $(\beta_c\rho^2/2\lsc^2)$
is equivalent to the ratio of rotational
kinetic energy density to magnetic pressure. It may therefore be thought of as a plasma `beta' for bulk rotation, 
rather than random ion motions. If
we consider the equatorial location of any given flux tube ($\rho=\rho_{0}$), then we find 
that the hot and cold plasma contributions to the
source function are equal at an equatorial radial distance $\rho_{T}$ given by:
\begin{align}
\rho_T^2 = 2 \chi\lsc^2 (\beta_h / \beta_c).
\label{eq:rho_trans}
\end{align}
 
 Beyond the transition distance $\rho_{T}$ the rotational kinetic energy of the plasma exceeds its thermal energy. 
 Therefore, centrifugal
 force (in a corotating frame) dominates pressure gradients
 for distances $\rho\gg\rho_{T}$ in determining both the magnetospheric current; 
 and the distortion to  the planetary dipole field required to maintain the magnetodisc's dynamic 
 equilibrium. Conversely, for $\rho\ll\rho_{T}$, the role
 of rotation is less important and plasma pressure determines disc structure. 
 Interestingly, $\rho_{T}$ may conceivably
 exceed the standoff distance of the dayside magnetopause under conditions
 where: 
 (i) hot plasma $\beta$ is very high compared to the cold plasma,
 (ii) plasma angular velocity is adequately low, or (iii) for a given temperature of cold plasma, 
 its density is small (such that the quantity $\lsc^2/\beta_c$ 
 becomes very large).
 We shall see in the following sections that average magnetospheric conditions at Saturn may yield transition distances at or inside the
 magnetopause (outer magnetosphere), while for Jupiter the magnetospheric current arises predominantly from hot pressure gradients.
 
 We shall now investigate the relatively simple expression for the magnetic potential of
 the homogeneous disc model. 
 We make use of the following equality, valid for dipolar magnetic field lines:
\begin{align}
\rho_{0} = r / (1-\mu^2).
\label{eq:dipole_line_shape}
\end{align}
\Eq{\ref{eq:dipole_line_shape}} indicates that dipolar field lines have parabolic shapes in the ($r,\mu$) 
co-ordinate plane. By definition, $\rho^2 = r^2 (1-\mu^2)$ and thus the argument of the exponential factor
in (\ref{eq:zdisc_g}) may be expressed in terms of $r$ and $\mu$. The full representation of the plasma source function
in these co-ordinates can be derived as (eliminating $\rho_{0}$ using \Eq{\ref{eq:dipole_line_shape}}):
\begin{align}
g_D(r,\mu) = \,\, & r^{-\chi} (1-\mu^2)^{\chi+1}  \notag \\
       & \left[\beta_{h} \chi (1{-}\mu^2)^2 + \frac{\beta_{c} r^2}{2\lsc^2} 
       \exp\left(-\frac{r^2}{2\lsc^2}\frac{1{-}(1{-}\mu^2)^3}{(1{-}\mu^2)^2}\right)   \right].   
\label{eq:g_zdisc_rmu}       
\end{align}

The source function for our simplified disc structure, \Eq{\ref{eq:g_zdisc_rmu}}, has been derived using
a dipolar magnetic field. Therefore it is an appropriate form for magnetic fields which may be decomposed into
the planetary dipole plus a small perturbation (in the sense that the perturbation is everywhere small compared to the dipole field strength).
In order to provide an approximation for this perturbation field due to the plasma disc, we shall 
calculate only the zeroth-order terms in the expansion for the magnetodisc potential described by
\Eq{\ref{eq:formal_mdisc_soln}}. By `zeroth-order', we mean the terms involving the
orthonormal basis function 
$(1-\mu^2) \jacpoly{0}{\mu} = (\sqrt{3}/2)(1-\mu^2)$, which involves the Jacobi polynomial
of order zero. This function represents 
the largest angular scale of the magnetodisc potential ($\mu^2$ dependence). The planetary dipole field
has this angular dependence and is thus included in the zeroth-order solution.

If we use \Eq{\ref{eq:formal_mdisc_soln}} to calculate 
the zeroth-order part of the potential, in conjunction with the explicit form of the source function for
our homogeneous disc.  \Eq{\ref{eq:g_zdisc_rmu}}, the resulting
expression is:
\begin{align}
\alpha_{0}(r,\mu) = \,\, & (1-\mu^2)/r \notag \\
                         & \left(  1 + \int_1^r \, u^2 g_0(u) \, du 
                         + r^3 \int_r^\infty \, u^{-1} g_{0}(u) \, du \right),  
\label{eq:zero_order_potl}
\end{align}
where we have defined $g_0$, the zeroth-order coefficient of the source function $g_{D}$, in terms of radial distance:
\begin{align}
g_{0}(r) = \frac{1}{4} \int_{-1}^1 \, g_{D}(r,\mu) \, d\mu.   
\label{eq:g0_defn}
\end{align}

The first factor in \Eq{\ref{eq:zero_order_potl}} is the unperturbed dipole potential, and
the integral terms in the second factor (enclosed by square brackets) represent the lowest-order
(largest angular scale)
perturbations due to the presence of the model plasma disc. It is evident from the integral limits
that any location which lies
outside the disc plasma will still experience a magnetic field due to all of the remote disc currents flowing
within the radial distance of such a point.  

\begin{figure}
\centering
\includegraphics[width=0.99\figwidth]{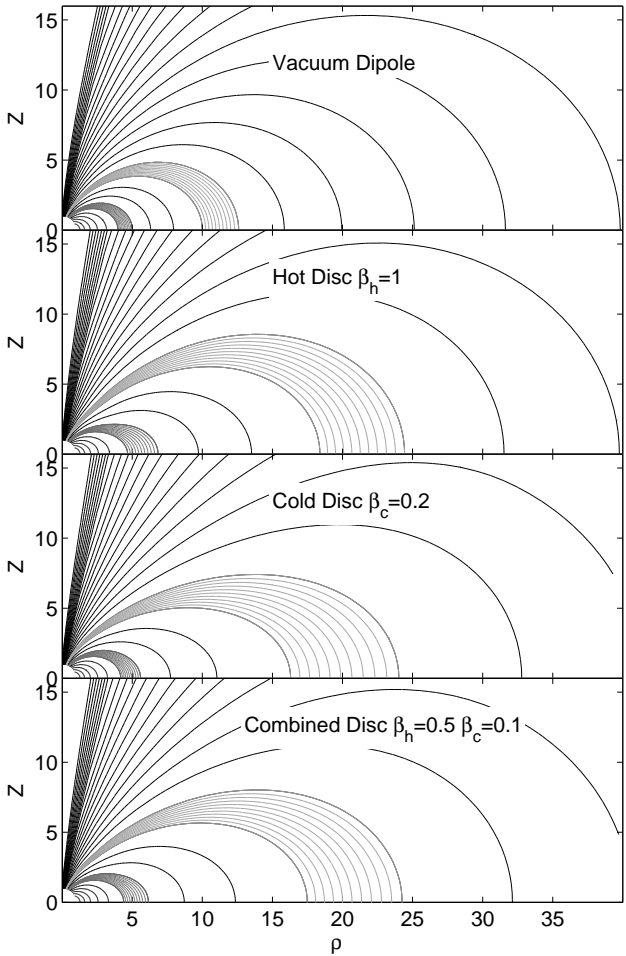}
\caption{Geometry of magnetic field lines for the zeroth-order homogeneous
disc models (see text). The solid black lines indicate contours which are 
spaced by uniform intervals in the logarithm of the magnetic potential. The dark grey
and light grey lines indicate regions which lie between the same values of magnetic
potential in each panel.}
\label{fig:homog_disc_field_lines}
\end{figure}

\Fig{\ref{fig:homog_disc_field_lines}} shows contours of
equal magnetic potential (i.e.\ magnetic field lines) for the function $\alpha_{0}$ evaluated for three
examples of the homogeneous disc model. The first model is a hot disc with $\beta_{h}=1$ and no
cold population ($\beta_{c}=0$); the second is a cold disc with $\beta_{h}=0$ and $\beta_{c}=0.2$;
and the third is a combined disc with $\beta_{h}=0.5$ and $\beta_{c}=0.1$. All the homogeneous discs
are assumed to be in perfect corotation with the planet, have uniform length scale $\lsc=1$ and uniform
field strength index $\chi=3$.
The disc models have had their structures truncated by setting their source function to zero for regions which 
are magnetically conjugate to equatorial distances $\rho_{0} < 5$ and $\rho_{0} > 35$.
The field lines of a vacuum dipole are also shown for comparison in the top panel. 
The closely-spaced dark grey and light grey field lines
indicate regions in each panel which cover the same intervals in magnetic potential. This may also be considered as an interval
in magnetic flux, since the total magnetic flux $\Phi_{B}$ threading the region from cylindrical
radial distance $\rho$ to infinity is a simple multiple of magnetic potential evaluated 
at this distance: $\Phi_{B}(\rho) = 2\pi\alpha(\rho, Z)$ (where the vertical co-ordinate $Z = r \cos\theta = r \mu$).
This relation follows from the definition of the magnetic field components in \Eq{\ref{eq:field_compts}} \citep{caudal}.
We see that the presence of the disc tends to `inflate' the dipolar magnetic field lines 
and shift them to larger equatorial crossing
distances. For the flux tubes highlighted, we see that this effect is quite pronounced for the dipolar field lines which cross
10--12 planet radii ($\RP$) at the equator (shaded light grey). These flux tubes are displaced to equatorial distances 
\unit[16\mbox{--}25]{\RP} by the disc models.
By contrast, the flux tubes shaded dark grey, situated near $\sim\unit[5]{\RP}$,  do not undergo as great a distortion in the presence of 
the disc. This part of the field may be considered as a `rigid', inner magnetosphere dominated by the internal planetary field.
We also note that the cold disc model produces outer magnetospheric field lines which are noticeably more oblate in shape compared to the other
discs: a field line crossing at a given equatorial distance does not rise as far above the equator in the cold disc model.
This property reflects the tendency of the cold plasma disc to be concentrated near the equatorial plane (according to the scale length
$\lsc$), thus producing stronger distortion 
in the near-equatorial segments of the model field.

\begin{figure}
\centering
\includegraphics[width=0.99\figwidth]{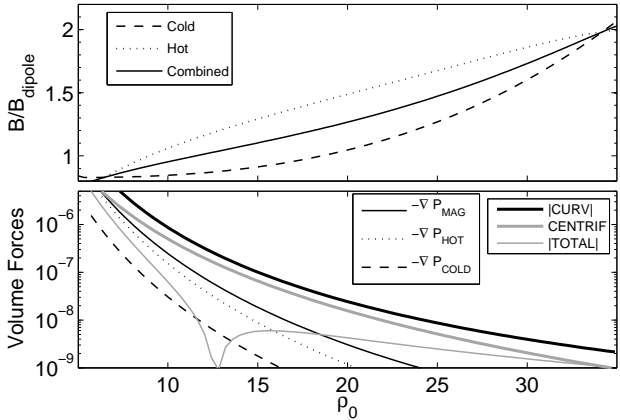}
\caption{Upper panel: field strength in the equatorial plane, relative to
that of a pure dipole field, as a function of
radial distance $\rho$, for various homogeneous disc models
(see legend and text). Lower panel: 
forces per unit volume (dimensionless) in the equatorial plane of the homogeneous, combined disc 
model, labelled according to line style.} 
\label{fig:homog_disc_balance}
\end{figure}

The outer magnetospheric flux tubes in the present example (field lines shaded light grey in 
\Fig{\ref{fig:homog_disc_field_lines}}) also become spread out over a larger radial distance
compared to the dipole configuration. 
In the upper panel of \Fig{\ref{fig:homog_disc_balance}}, we compare the 
corresponding equatorial profile of field strength between the disc models and the dipole field. For all disc models, the ratio
of total to dipole field strength increases monotonically with distance, indicating that the
magnetodisc field is more uniform than the dipole. We also see that each model has a characteristic
distance which separates an inner region where the field strength ratio $B/\Bdip<1$ from an outer region where this
ratio exceeds unity. This feature is observed in actual planetary ring currents, and arises from the finite extent of the current 
region and the solenoid-like nature of the corresponding magnetic field
(e.g.\ \citet{sozou1970} and Figures~1, 4 and 5 of \citet{arridge2008}). 
Near the inner boundary of the 
equatorial disc current, the vertical magnetic field generated by this current alone opposes the
planetary field, while the opposite is true near the outer disc boundary, where the disc field enhances that of the planetary dipole. 
We thus expect and find that the ratio of total magnetic field to that of a pure dipole monotonically increases
from values less than unity near the inner edge of the disc ($\sim\unit[5]{\RP}$) to values larger than unity near
the outer edge  ($\sim\unit[35]{\RP}$).

We now consider the radial profiles of the main forces involved
in the dynamic equilibrium of the homogeneous disc structure. We do not expect these 
forces to be in perfect balance for our zeroth-order model,
since this is only one component of the many required to retrieve the full solution, and corresponds to
the largest angular scales of the problem ($\sin^{2}\theta$ dependence). We plot the equatorial centrifugal force, 
magnetic forces and 
plasma pressure gradients as a function of distance in the lower panel
of \Fig{\ref{fig:homog_disc_balance}} for the homogeneous, combined disc model described above.
The total magnetic force ($\vec{J}\times\vec{B}$)
is the sum of the
magnetic pressure gradient and the curvature force. When these two components have equal magnitude and
opposite sign, the total magnetic force is zero. In the figure, the absolute value of the negative
curvature force is displayed. Thus it is the vertical difference between this curve and that for the
magnetic pressure gradient which indicates the magnitude of the total magnetic force.
The plots show us that, for the outer region where the ratio $B/\Bdip>1$ and $\rho>\unit[12]{\RP}$,
the force of highest magnitude is that due to magnetic curvature, followed
by centrifugal force (factor of $\lesssim2$ smaller than
curvature force) and magnetic pressure gradient (factor of $\gtrsim5$ smaller than curvature force). The hot plasma pressure 
has a gradient about half the magnitude or less of that for magnetic pressure, while the weakest force in this model is the gradient due to
cold plasma pressure. We see that the sum total of the forces is less than
one \unit{per\,cent} of the curvature force at $\rho=\unit[12]{\RP}$,
and less than ten \unit{per\,cent} for $\rho=\unit[10\mbox{--}16]{\RP}$. This aspect of the total force profile is an indication of the degree to which
dynamic equilibrium is maintained within the zeroth-order component of the full solution.

The transition distance for the combined disc model is $\rho_T\sim\unit[5.5]{\RP}$, using 
\Eq{\ref{eq:rho_trans}}. It is also evident from \Fig{\ref{fig:homog_disc_balance}} 
that, for distances much greater than this value, we are by definition in the part of the magnetosphere where centrifugal force
dominates hot plasma pressure, and where the force balance principally involves the centrifugal and magnetic curvature forces.
As we approach the transition distance from the outer disc, the other forces due to plasma pressure and magnetic pressure become
comparable to curvature force; and thus become more significant in determining stress balance and disc structure.

\subsection{Model Inputs and Boundary Conditions}
\label{sec:modelinputs}
In this section, we summarise the model inputs we have used to determine the equatorial boundary conditions
for our calculations of the Kronian magnetodisc field. These inputs have been drawn from a variety of observational
studies using the plasma instruments aboard the \Cassini spacecraft. They fall into the following 
four categories, each described in 
more detail in the subsections below: (i) composition, (ii) temperature and density, 
(iii) hot plasma pressure; and 
(iv) rotation. 

\subsubsection{Equatorial Plasma Composition}
\label{sec:plasmacompn}
The equatorial composition of the cold plasma at Saturn is required in the model for the computation of the 
scale length $\lsc$ (\Eq{\ref{eq:ldefn}}), assumed constant along magnetic field lines. The composition
of the plasma is determined by the relative densities of water group ions (mass $m_{W} = \unit[18]{amu}$) and 
protons (mass $m_{P} = \unit[1]{amu}$). Following \citet{caudal} we represent the disc ions as having a mean
mass $m_{i}$ between these two limits, given by:
\begin{align}
m_i = \frac{n_Wm_W + n_Pm_P}{n_W + n_P},  
\label{eq:mi_defn}
\end{align}
where the symbol $n$ denotes number density, with obvious subscripts indicating water group
and proton components. 

\begin{figure}
\centering
\includegraphics[width=0.99\figwidth]{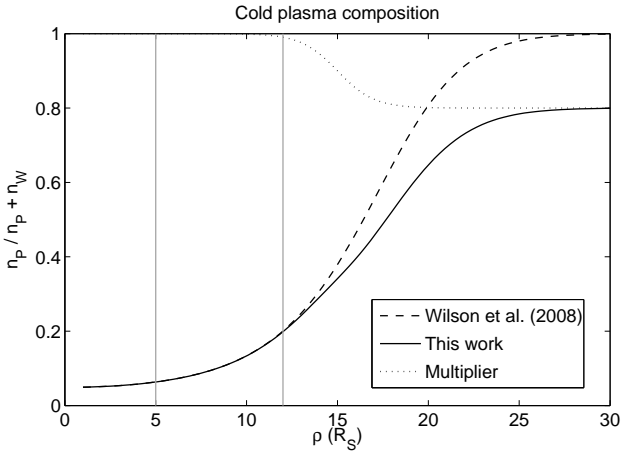}
\caption{
Profiles of proton number fraction in the cold disc plasma. Dashed line: profile determined from the fits to plasma density moments
by \citet{wilson2008}. Solid line: profile used in this work. Dotted line: the multiplying function used to constrain the outer
magnetospheric composition (see text). Vertical lines indicate the range of validity of the fitted functions of \citet{wilson2008}.
} 
\label{fig:plasmacompn}
\end{figure}

In order to capture the behaviour of $m_{i}$ as a function of radial distance,
we employed the formulae of \citet{wilson2008}, who determined and fitted density moments for water group ions
and protons using observations by \CAPS. 
\citet{wilson2008}'s observations
sampled five equatorial spacecraft orbits in the distance range $\sim\unit[5.5\mbox{--}11]{\RS}$
between October 2005 and April 2006. The orbits were chosen as mission segments during which \CAPS
ion mass spectrometer obtained sufficient coverage of the plasma particle distribution to allow reliable computation of moments.
For the purposes of our modelling, we used the Gaussian fits to water ion density and proton density
by  \citet{wilson2008} to compute the following number fraction of protons as a function of radial distance:
\begin{align}
\frac{n_H} {n_W + n_H} = & \, \frac{f_M(\rho)}
{1 + \left(A_W/A_H\right) \exp\left[(B_H-B_W) \rho^2\right]}   \notag \\
f_M(\rho) = & \, 0.1 \left[1-\tanh\left(\frac{\rho-15}{2}\right)\right] + 0.8 , 
\label{eq:nhfrac_profile}
\end{align}
where the function parameters provided by \citet{wilson2008} are 
$A_W=\unit[161.5]{cm^{-3}}$, $A_H=\unit[8.3]{cm^{-3}}$,
$B_W=\unit[0.042]{\RS^{-2}}$ and $B_H=\unit[0.031]{\RS^{-2}}$.
Since these fitted functions are based on observations in the distance range \unit[5\mbox{--}12]{\RS}, we 
used the hyperbolic tangent function $f_{M}$ to place the  additional constraint that the proton number fraction approaches
\unit[80]{per\,cent} in the outer magnetosphere. This plasma composition was determined by \citet{arridge2007} to provide good agreement
with both electron densities observed by \Cassini in the outer magnetosphere and the surface mass density of the Kronian plasma disc.
The latter quantity was deduced in the same study from the analysis of magnetic signatures of transient excursions by the spacecraft into
the magnetodisc current sheet. 
\Fig{\ref{fig:plasmacompn}} 
shows plots of the fitted composition profile from \citet{wilson2008};
the extrapolation of this profile beyond the range of validity (\unit[5\mbox{--}12]{\RS}); the profile used in the current work; and the
multiplier function $f_{M}$.

\subsubsection{Equatorial Plasma Temperature and Density}
\label{sec:coldplasma}
\begin{figure}
\centering
\includegraphics[width=0.99\figwidth]{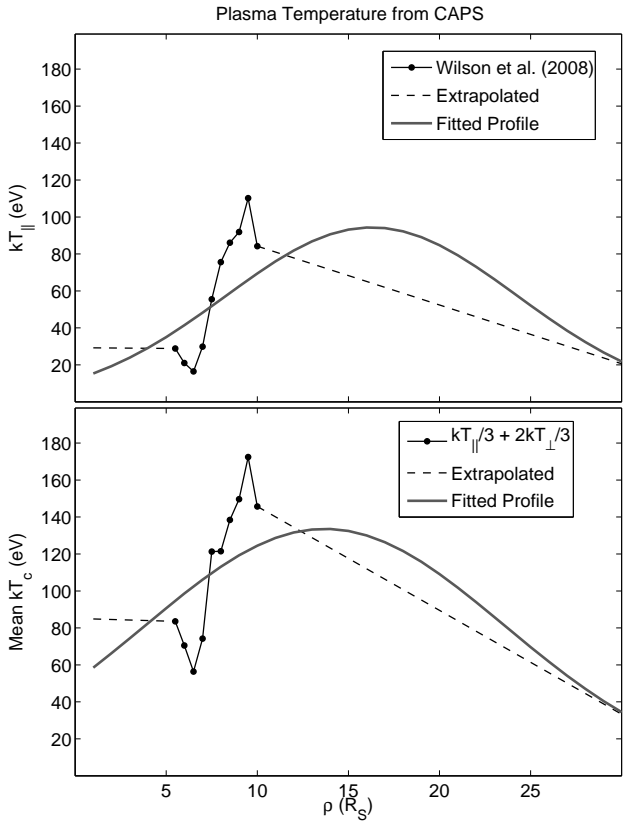}
\caption{
Profiles of parallel (top panel) and mean (lower panel) total temperature for the cold disc plasma. 
Solid lines with dots: total plasma temperatures derived from the moments for water group ions and protons, determined from \Cassini 
plasma data by \citet{wilson2008} (see text). Dashed lines: extrapolated temperature profiles, derived by assuming that the
individual ion species have temperatures equal to those at the nearest location in the tabulation of  \citet{wilson2008}.
Solid grey lines: the final model inputs, obtained by second-order polynomial fits to the temperature profiles derived from the data.
} 
\label{fig:plasmatemp}
\end{figure}

For the cold plasma population, the contribution to the source function $g$ (\Eq{\ref{eq:caudal_de}}) takes the form
(see also \Eq{\ref{eq:zdisc_g}}):
\begin{align}
g_c(r,\mu,\alpha) = 
\rho^2\exp\left(\frac{\rho^2-\rho_0^2}{2\lsc^2}\right)\left(\deriv{P_{c0}}{\alpha} 
+ \frac{P_{c0}}{\lsc^2\Bteq}\right),         
\label{eq:cold_g}
\end{align}
where the geometry of the magnetic field, represented by $\alpha$, determines the mapping between $\rho$ and $\rho_{0}$ along
a field line, as well as the equatorial field strength  $\Bteq$. In
order to compute the scale length $\lsc$ in the equatorial plane,
we require the equatorial distribution of plasma temperature. Strictly speaking the latter should be the field-parallel temperature, since $\lsc$ is
associated with force balance parallel to the magnetic field. In addition, a knowledge of both plasma temperature and density is required to specify the
equatorial pressure $P_{c0}$. 
To satisfy these requirements, we appealed to the study by \citet{wilson2008} which 
provided tabulated measurements of both parallel and perpendicular temperature for the thermal water group ions and protons
at Saturn. These tabulations contain average temperatures over intervals of
radial width \unit[0.5]{\RS} near  the planet's equatorial plane,
extending between radial distances \unit[5.5\mbox{--}10]{\RS}.
To obtain total plasma temperatures for modelling purposes, we began by combining the tabulated ion and proton temperature values from 
\citet{wilson2008} as follows: 
\begin{align}
T_\| & = \frac{n_W T_{W\|} + n_H T_{H\|}}{n_W + n_H}, \notag \\
T_\perp & = \frac{n_W T_{W\perp} + n_H T_{H\perp}}{n_W + n_H}, \notag \\
T_c & = \frac{T_\| + 2T_\perp}{3},    
\label{eq:plasmaT_defn}
\end{align}
where $n_{W}$ and $n_{H}$ are the respective water-group ion and proton 
number densities from \citet{wilson2008} (see \S\ref{sec:plasmacompn}); the symbol $T$ represents temperatures;
and the subscripts $\|$ and $\perp$ are associated with thermal motions parallel and perpendicular to the magnetic field.
The quantities $T_{\|}$ and $T_\perp$ are average parallel and perpendicular temperatures for the cold plasma (weighted by number density between protons and water-group ions),
while $T_{c}$ is an
appropriately weighted mean. In \Fig{\ref{fig:plasmatemp}} we show plots of the 
radial profiles
of $\kb T_\|$ and $\kb T_c$, expressed in units of \unit{eV}. 

While the data provided by \citet{wilson2008} are valuable for our work, we still need to
assign temperatures to those regions of the magnetosphere outside the reach
of this study, i.e.\ $\rho_0 < \unit[5]{\RS}$
and $\rho_0 > \unit[10]{\RS}$. In order to do this, we have assumed that the 
individual water group ion and proton temperatures in these regions
are equal to those measured by \citet{wilson2008} at the closest relevant points (i.e.\ at \unit[5.5]{\RS} and \unit[10]{\RS}
respectively). 
We then compute extrapolated total temperatures using the weighted sum (according to
plasma composition) given by
\Eq{\ref{eq:plasmaT_defn}}. Even though the individual temperatures of the heavy ions
and protons are assumed fixed in this extrapolation, the variation in the plasma composition
produces a total temperature which steadily decreases with distance beyond \unit[10]{\RS}.
This behaviour is due to
the increasing fraction of the relatively cold protons in the plasma in the more distant magnetosphere.
We show the extrapolated total parallel and mean plasma temperature as dashed lines 
in \Fig{\ref{fig:plasmatemp}}. To obtain the final, realistic input
temperature profiles for the cold plasma, we applied second-order polynomial fits to the 
data-derived profiles of $T_{\|}$ and $T_{c}$.
We believe that such an approach is justified in light of the fact that the observations by \citet{wilson2008} show variability in
temperature moments within their \unit[0.5]{\RS} bins, typically by factors between two and five, even for data from the same orbit.
The final fitted profiles are shown as grey curves in \Fig{\ref{fig:plasmatemp}}.
We note that our fitted values for $T_{\|}$ in the range \unit[10\mbox{--}15]{\RS} are somewhat
lower than the value of a few hundred \unit{eV} corresponding to the observations of \citet{mcandrews2009}.
However, the large-scale trend of our fit agrees with these data and our fitted temperatures are of similar order
of magnitude. We aim to incorporate further plasma temperature measurements as they become available.

We used the profiles of $T_\|$ in conjunction with our mean ion mass $m_i$ (\S\ref{sec:plasmacompn}) to 
compute the plasma scale length. We used the  $T_c$ profiles in order to compute the cold plasma pressure in the equatorial
plane, according to the following formula, adapted to dimensionless form from \citet{caudal}:
\begin{align}
P_0(\alpha) = 2 N_L(\alpha) (\kb T_c)^{*} / V_W(\alpha).
\label{eq:coldpressdefn}
\end{align}
Here, the dependence upon the local value of $\alpha$ (i.e.\ the particular field line)
is indicated for the dimensionless quantities $V_W$ and $N_L$; these are, respectively, the
weighted unit flux tube volume and the flux tube content. Considering these two 
quantities for the moment, their definition 
is based on the usual concept of the unit flux tube volume, which we define in
our normalised system as:
\begin{align}
V(\alpha) = \int_0^{s_B}   \, ds / B,
\label{eq:ftvol_defn}
\end{align}
where the integral is taken along a magnetic field line of length $s_B$ between its southern
and northern ionospheric footpoints; $ds$ is an element of arc length along the field line; and $B$
is local field strength. Given the relation between equatorial field strength and the increment in magnetic potential
(see derivation of \Eq{\ref{eq:zdisc_g_trans}}), it follows that $2\pi
V(\alpha)\abs{d\alpha}$ represents the normalised volume 
between two magnetic shells corresponding to the interval \mbox{$[\alpha,\alpha+d\alpha]$}.
This same volume is threaded by an increment $2\pi\abs{d\alpha}$ 
of normalised magnetic flux.

Using this definition of the unit flux tube volume, we can construct the weighted flux tube volume as follows:
\begin{align}
V_W(\alpha) = \int_0^{s_B} \, \exp\left(\frac{\rho^{2}-\rho_{0}^{2}}{2\lsc^2}\right)\, ds / B,
\label{eq:wftvol_defn}
\end{align}
where the exponential weighting factor is a consequence of field-aligned pressure balance for the cold rotating
plasma (see also \Eq{\ref{eq:zdisc_g_trans}}). The flux tube content $N_L$ is defined as the
number of cold ions per unit of magnetic flux. That is, the quantity $2\pi N_L(\alpha)\abs{d\alpha}$
is the number of ions within the volume bounded by the magnetic shells corresponding to the interval
$[\alpha,\alpha+d\alpha]$ in magnetic potential.
The factor $(\kb T_{c})^{*}$ in \Eq{\ref{eq:coldpressdefn}} is the dimensionless form
of the thermal energy corresponding to the averaged plasma temperature $T_c$
defined in \Eq{\ref{eq:plasmaT_defn}}. We obtain this factor 
through division by the energy scaling factor in \Table{\ref{scales_appendix_table}}.

\begin{figure}
\centering
\includegraphics[width=0.99\figwidth]{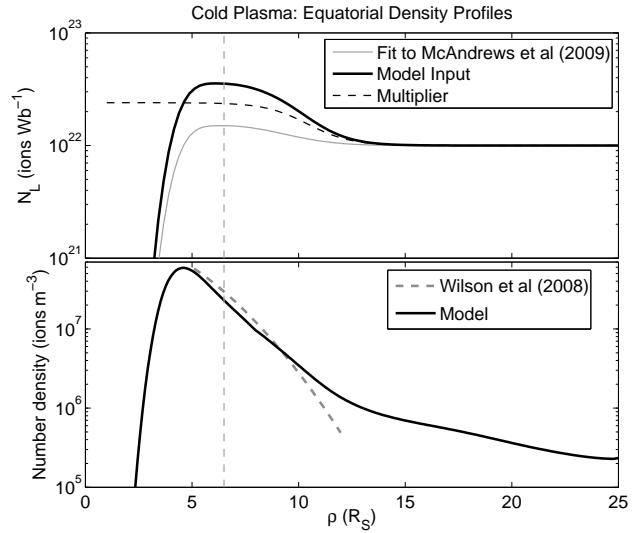}
\caption{
Profiles of cold plasma flux tube content (top panel) and number density (lower panel) 
in Saturn's equatorial plane. The top panel shows: radial profiles of flux tube content $N_L$ in ions per Weber
for the Gaussian fits to the original observations of \citet{mcandrews2009} (grey line); the shape of the
multiplying function used to modify this profile (dashed line, see text); and the final form of the profile used
for the disc models (black line). The vertical line shows the division between regions where different Gaussian
profiles were used to fit the $N_{L}$ data. The lower panel shows the agreement between
the ion number density profile from the 
observational fits of
\citet{wilson2008} with that derived from a disc model with magnetopause radius 
$\RMP = \unit[25]{\RS}$. 
} 
\label{fig:plasmadens}
\end{figure}

By specifying a profile of flux tube content in the Caudalian model rather than density, it is more straightforward
to mimic realistic changes associated with a plasma which is `frozen-in' to the magnetospheric field. Our profile
for the flux tube content was obtained by fitting estimates of this quantity from the
work by \citet{mcandrews2009}, extended to cover inner
magnetospheric regions \unit[4\mbox{--}10]{\RS} (H.~J.\ McAndrews, private communication).
These authors used \Cassini nightside near-equatorial plasma observations by \CAPS 
in conjunction with the magnetospheric field model by 
\citet{khurana2006} in order to estimate $N_{L}$ through  
a force balance relation similar to \Eq{\ref{eq:coldpressdefn}}. 
We have fitted the flux tube content measurements with two Gaussian
functions, constrained to meet continuously at \unit[6.5]{\RS}.
The entire fitted profile for $N_{L}$ used in the model is shown
in \Fig{\ref{fig:plasmadens}}.
In order to bring this profile of flux tube content 
into reasonable agreement with the density moments
provided by \citet{wilson2008} (who studied orbits distinct from those
used by \citet{mcandrews2009}), we multiplied them
by a smooth correction function whose shape, but not absolute scale,
is also shown in \Fig{\ref{fig:plasmadens}}.

\subsubsection{Hot Plasma Pressure}
\label{sec:hotplasmapressure}
\citet{caudal} assumed that the hot magnetospheric plasmas filled each flux tube such that
each flux tube can be characterised by a particular equatorial pressure $P_{h0}$ and volume per unit flux $V$, referred
to as the hot plasma approximation. Using the ideal gas equation per unit flux one can show that the
product $P_{h0}V$ is equal to $N_0\kb T_h$ where $N_0$ is the number of ions per unit flux and $\kb T_h$ is the mean kinetic
energy of the ions (see also \S\ref{sec:coldplasma}). \citet{caudal} used observations from the Jovian magnetosphere to show that $\kb T_h$ did not vary
appreciably with $L$. In the absence of plasma sources the time-stationary
radial (cross-$L$) transport of plasma can be described using a one
dimensional diffusion equation. \citet[][and references therein]{caudal}
showed that the plasma tended to have a uniform distribution in $L$ when the
rate of loss of particles due to pitch angle scattering into the loss cone
was negligible compared to the rates for cross-$L$ transport. This reasoning
led \citet{caudal} to conclude that $P_{h0}V$ and $N_0(L)$ were independent
of $L$ and hence that under rapid radial diffusion the hot plasma in the
Jovian magnetosphere behaves isothermally rather than adiabatically:
$P_{h0}V^\gamma=\mathrm{const}$ where $\gamma=1$. \citet{caudal} used
published energetic particle pressures and magnetic field models in order 
to show that the particles did indeed behave isothermally beyond
$\sim\unit[18]{\RJ}$, but adiabatic on smaller $L$-shells. In further work, \citet{caudalconnerney1989} made $\gamma$ a free parameter in a fit of the model to Voyager magnetometer data. They found that $\gamma=0.88$ beyond \unit[9]{\RJ}, suggesting the presence of non-adiabatic cooling processes during inward diffusion, losses, and violations of the first and second adiabatic invariants.

Following \citet{caudal}, we parametrised the distribution of hot plasma 
pressure in our model by appealing to observations, using the same $P_{h0}V^\gamma=\mathrm{const}$ theoretical framework. The data required were taken from the study by \citet{sergis2007}, who determined pressure moments
for ions with energy $>\unit[3]{keV}$ from the measurements of \MIMI.
The observations presented by these authors were acquired
within the distance interval \mbox{$5<\rho<20 \,\unit{\RS}$}
over eleven consecutive near-equatorial orbits of the spacecraft between
late~2005 and early~2006. An important result to emerge was that, within
this `hot population', particles with energy $>\unit[10]{keV}$ carried half
of the total pressure, but contributed only $\sim\unit[10]{per\,cent}$ of the total number density. We shall see in the later sections that the hot plasma pressure for typical conditions at Saturn may exceed that of the colder population (see \S\ref{sec:coldplasma}) by up to an order of magnitude; it was therefore important to include a representation of this hot pressure component in our magnetodisc model's source function.

\begin{figure}
\centering
\subfigure[][]{
\label{fig:input_hot_press_a}
\includegraphics[width=0.99\figwidth]{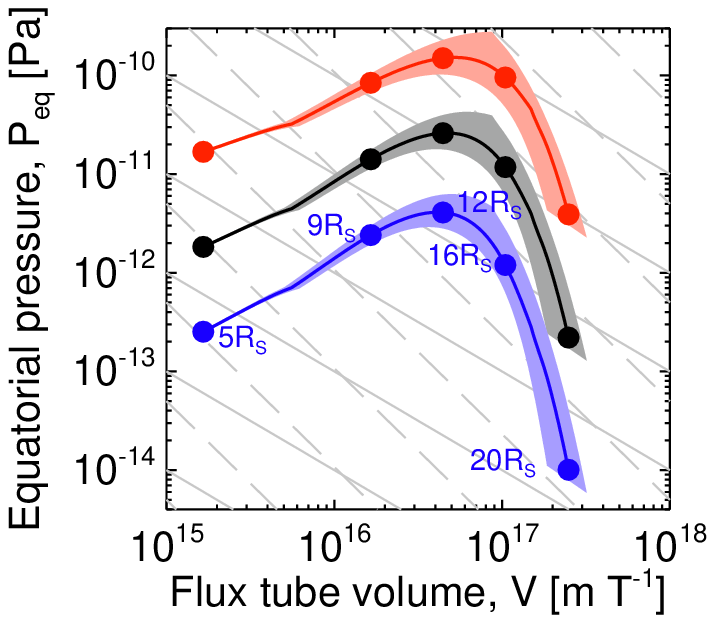}}\\[-0.3cm]
\subfigure[][]{
\includegraphics[width=0.99\figwidth]{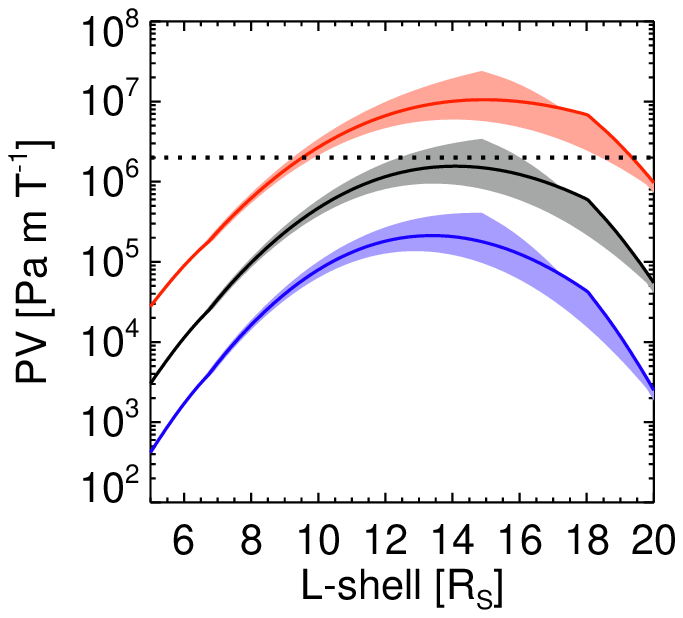}}
\label{fig:input_hot_press_b}
\caption{Hot plasma pressure $P_{h0}$ as a function of unit flux tube volume $V(\alpha)$ in 
SI 
units (panel a) and the product of hot plasma pressure and unit flux tube volume as a function of 
$L$ (panel b) in Saturn's equatorial plane. The coloured curves employ the fitted pressure profiles 
determined by \citet{sergis2007} to the highly variable pressure data from \MIMI. 
The colours represent the disturbed (red), average (black) and quiescent (blue) ring current states, 
in the parlance of these authors. The solid coloured curves represent the profiles for a nominal 
magnetopause standoff distance of \unit[25]{\RS} and the shaded regions represent the 
variability introduced into these profiles due to the changing upstream solar wind dynamic pressure 
affecting the global magnetic field configuration. In panel (a) the solid lines represent isotherms and 
the dashed lines are adiabats. Particular $L$-shells are indicated by the solid dots and annotation. In 
panel (b) the horizontal 
dotted line shows the profile corresponding to the constant value of 
$K_h = PV = \unit[2\cdot10^6]{Pa\,m\,T^{-1}}$ (see text). This value may be compared
with the data-derived curves, and we use this profile as a simple means of representing
`average conditions' within the highly-variable pressure distribution beyond $\rho=\unit[8]{\RS}$.}
\label{fig:input_hot_press}
\end{figure}

We used an empirical magnetic field model to determine the unit flux tube
volume $V(\alpha)$ as a function of $\rho$. The empirical model comprised an
un-tilted dipole and CAN current sheet, where the parameters of the model
current sheet were dependent on the distance to the subsolar magnetopause
\citep{bunce2007}. We note in passing that the results of this analysis are
not significantly altered by using an alternative model, such as
\citet{khurana2006}. The second-order fits to plasma $\beta$ as a function
of $L$ by \citet{sergis2007} were then used to provide values of the hot
plasma pressure $P_{h0}$ at the equator, using the model magnetic field
strength to calculate pressure from plasma $\beta$. The results are
presented in \Fig{\ref{fig:input_hot_press}}a, where we show $P_{h0}$ as a
function of $V(\alpha)$ for the three different fits to the highly-variable
hot pressure data presented by \citet{sergis2007}, which they referred to
(in order of increasing plasma $\beta$) as the quiescent (blue), average
(black) and disturbed (red) ring current. The shaded regions indicate the
variability introduced by modifying the subsolar standoff distance of the
magnetopause which modifies the parameters of the CAN model
\citep{bunce2007}. By comparison between the calculations and the isotherms
(solid) and adiabats (dashed) one can see that there is only a very narrow
region in $L\sim\unit[12\mbox{--}16]{\RS}$ where the transport can be
considered to be either isothermal or adiabatic. Inside \unit[9]{\RS}
Saturn's neutral \chem{OH} cloud is a strong absorber of energetic particles
and losses might reasonably account for the decrease in pressure at smaller
values of $L$ (and $V(\alpha)$). In support of this, it is known that the hot oxygen temperature is approximately 
constant with $L$ \citep{dialynas2009} and that the hot oxygen contributes the most to the hot pressure $P_{h0}$.  
Hence, a reduction in pressure is related to a decrease in the number of hot oxygen ions per unit flux. At larger distances the 
pressure varies more steeply than $P_{h0} V(\alpha)^{5/3}$ suggesting a reduction in pressure.  This may be related to 
the observed warping of the magnetic equator \citep{arridgewarp2008} which implies that particle pressures measured in 
the rotational equator will be smaller than at the magnetic equatorial plane. Energetic particle pressures beyond 
\unit[20]{\RS} presented by \citet{sergis2009} support the fact that the pressure appears to be underestimated 
by the fits in \citet{sergis2007}.

The product $P_{h0} V(\alpha)$ was also determined as a function of $\rho$
and shown in figure \Fig{\ref{fig:input_hot_press}}b. As expected, $P_{h0}
V(\alpha)$ increases linearly with $L$ within, and just beyond Saturn's
neutral cloud, due to the increasing flux tube volume and pressure and peaks
near \unit[13\mbox{--}15]{\RS} before falling with increasing $L$. Using the
pressures published by \citet{sergis2009} and the calculated flux tube
volumes beyond around \unit[16]{\RS} the value of $P_{h0} V(\alpha)$ for
$L>\unit[16]{\RS}$ is greater than $\sim\unit[5\cdot10^5]{Pa\,m\,T^{-1}}$.
Nevertheless, the entirety of the $P_{h0} V(\alpha)$ profiles reflects the
strong variability in hot plasma $\beta$ for Saturn's magnetosphere. It is
important to note in this context that the different ring current `states'
for plasma $\beta$ fitted by \citet{sergis2007} represent the range of values of this parameter over several different orbits and magnetospheric configurations, as well as a wide range of radial distances and local times within each orbit.

In view of the strong variability in this parameter and its general decline with decreasing distance inside 
$\sim\unit[10]{\RS}$, we adopted a simple representation for its global behaviour, similar to that 
of \citet{caudal}. We composed a profile of $P_{h0}$ by setting the product $P_{h0}V(\alpha)$ 
to a constant value $K_{h}$ beyond $\rho=\unit[8]{\RS}$; and by decreasing hot pressure linearly 
with decreasing $\rho$ inside this distance, according to the  formula $P_{h0}(\rho) = P_{h0}(\unit[8]{\RS}) 
\times (\rho / 8)$. We then retrieved $P_{h0}$ values in our model's outer magnetosphere beyond 
$\unit[8]{\RS}$ through $P_{h0} = K_{h} / V(\alpha)$. This form gives a more realistic response 
of the value of hot pressure to different configurations of the outer magnetosphere (from expanded to compressed) 
than would a single function of $\rho$ alone. In addition, the parameter $K_{h}$ gives a compact 
representation of the `level of activity' of the ring current in the
model, and also reduces the number of free parameters. 
We intend to pursue a future parametric study of disc structure dependent on this 
parameter and magnetopause radius. For the purpose of this introductory study, we 
set the value $K_{h} = \unit[2 \cdot 10^{6}]{Pa\,m\,T^{-1}}$ in our calculations 
(the scaling factor for normalised $K_{h}$ is given in \Table{\ref{scales_appendix_table}}). 
This value according  to \Fig{\ref{fig:input_hot_press}} represents a ring current somewhat more 
`disturbed' than the average state. 

\subsubsection{Plasma Rotation}
\label{sec:plasmarotn}
\Eq{\ref{eq:zdisc_g}} for the plasma source function includes the scale length $\lsc$, itself dependent on the 
angular velocity of the cold rotating plasma at each point along the equatorial plane. Under the steady-state 
assumption of the model, this angular velocity $\omega$ is constant along a magnetic field line, i.e.\
$\omega$ is expressible as a function of $\alpha$ alone. Thus it is observations of plasma $\omega$
which we seek in order to complete the equatorial boundary conditions for our model calculations.
To construct a model for the azimuthal velocity $v_\phi$ of the plasma we used data from studies by 
\citet{kane2008} and \cite{wilson2008}. The study by \citet{kane2008} provided measurements 
of $v_\phi$ through analysis of ion velocity anisotropies, acquired in Saturn's outer magnetosphere 
by \INCA, a detector of \MIMI. 
These data were acquired from the ion mode of the \INCA instrument.
As well as the estimated uncertainty of $\sim\unit[20]{per\,cent}$ 
in their individual velocity measurements, \citet{kane2008}'s results also show considerable variability, 
around factors of two, within subsets of their measurements acquired near the same radial distance. This variability 
is attributable to the underlying set of spacecraft orbits sampling different local times and magnetospheric 
configurations (e.g.\ the \INCA data used were obtained in the dawn sector for $\rho<\unit[25]{\RS}$ 
and in the midnight sector outside this distance). Following a different approach with a different data set, 
\citet{wilson2008} determined $v_\phi$ by fitting drifting bi-Maxwellian velocity distributions to 
\CAPS ion mass spectrometer data. They presented quadratic fits of $v_\phi(L)$ for the region between 
$5.5$ and $\sim\unit[10]{\RS}$.

\begin{figure}
\centering
\includegraphics[width=0.99\figwidth]{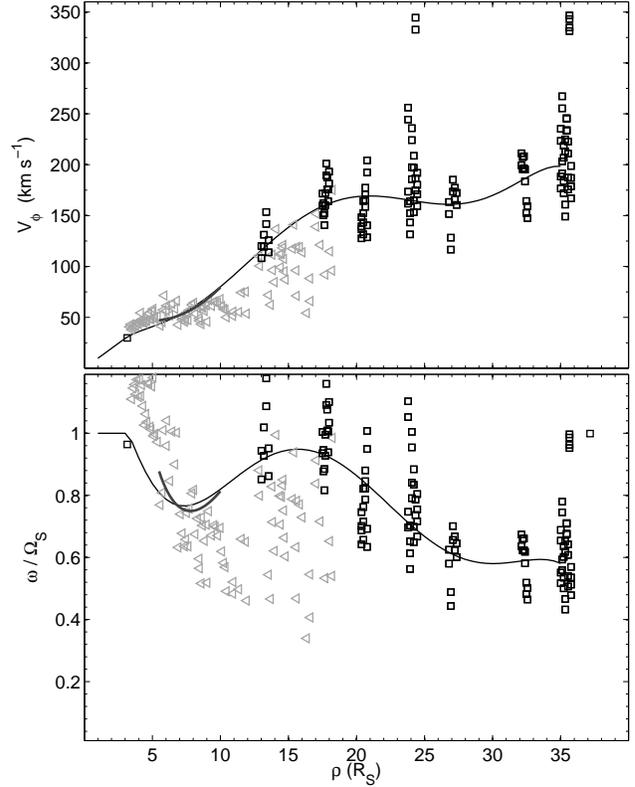}
\caption{Upper panel: A polynomial fit of order six for the azimuthal plasma velocity
(thin solid curve) compared with observations of the plasma azimuthal velocity in
Saturn's magnetosphere. The squares are data from \citet{kane2008} and were
used for our fit. The thick, dark grey line is 
\citet{wilson2008}'s empirical profile derived from their data,
which were also used for our fitting. The light triangles were
derived from \Voyager data by 
\citet{richardson1998} and are shown for comparison only but were not used in our fit.
Lower panel: The same comparison in the upper panel is shown with the azimuthal velocity
transformed to angular velocity in the models and observations.}
\label{fig:plasma_omega}
\end{figure}

While a fully self-consistent model would include the influence of magnetospheric configuration on the profile of $v_\phi$ and angular 
velocity $\omega$, we shall address this issue in a future study. 
For the present purpose, we use a profile of $v_\phi$ versus $\rho$ obtained by fitting a 
sixth-order polynomial to points from the model of \citet{wilson2008} and points taken from 
Figure~4 of \citet{kane2008}. 
Inside of \unit[3.1414]{\RS} we assumed the plasma is in ideal corotation with an angular velocity of 
\unit[1.638\cdot10^{-4}]{rad\,s^{-1}} (a period of \unit[10.65]{h}).
For the purposes of fitting, 
we used a constant value of 
$v_\phi$ outside \unit[25]{\RS} equal to the average value of the outer magnetospheric
observations from \citet{kane2008} and \citet{mcandrews2009}.
We found that this approach produced a well-behaved fit in the outer
magnetosphere without large `oscillations', as well as good agreement in the inner magnetosphere
with the data of \citet{wilson2008}. For the calculations in this 
paper, we used the resulting polynomial fit to represent plasma angular velocity
throughout the modelled magnetosphere. 
However we emphasise that fixing the value of $v_\phi$ to \unit[169.25]{km\,s^{-1}} beyond
\unit[25]{\RS} 
does not significantly alter the conclusions of our study.
The $v_\phi$ and $\omega$ profile corresponding to this 
fit are illustrated in \Fig{\ref{fig:plasma_omega}}. 
\comment{Note that whilst the modelled velocity is slightly 
smaller than the mean velocities presented by \citet{kane2008} in the outer magnetosphere, plasma velocities presented 
by \citet{mcandrews2009} are slightly lower than \citet{kane2008} and so the model is fully consistent 
with the published azimuthal velocities acquired by \Cassini.} 
For further comparison, we include the \Voyager velocity
measurements by \citet{richardson1998} in the figure,
but emphasise that we did not use these measurements to derive our
fitted profiles. At distances smaller than $\sim\unit[10]{\RS}$, the model curve
agrees well with the \Voyager data. Beyond this distance, the model has values higher than the mean \Voyager values,
but is still consistent with the full range of these measurements. 

For the information of other modellers, we also present here the 
seven-element vector \vec{C} of polynomial coefficients for the fitted plasma velocity profile.
The following coefficients generate $v_\phi$ in \unit{km\,s^{-1}}:
\begin{align}
v_\phi(\rho) =     &\sum_{n=0}^{6} C_{n} \rho^{n},
\qquad\rho\geq\unit[3.1414]{\RS}, \notag \\
C_{0} = & -15.09,\qquad   C_{1} = 28.16, \notag \\
C_{2} = & -6.359,\qquad   C_{3} = 0.7826, \notag \\
C_{4} = & -0.043,\qquad  C_{5}= 1.065\cdot10^{-3}, \notag \\
C_{6} = & -9.762\cdot10^{-6}.   
\label{eq:omega_prof}
\end{align}

\section{Magnetodisc Models}
\label{sec:saturnmodel}
Having described our methods for incorporating equatorial observations of plasma properties into the
Caudalian model formalism, we now turn our attention to some example model outputs and how such
calculations may be used to infer some important physical aspects of magnetodisc structure at Saturn.
We shall firstly consider some aspects of force balance in a disc formed under average solar
wind pressure conditions and magnetopause size as observed in the \Cassini era.

\subsection{Magnetodisc Structure for Average Magnetopause Size}
\label{sec:avg_mdisc}
The probability distribution of magnetopause
standoff distance at Saturn was determined by \citet{achilleos2008} who surveyed
magnetopause crossings of the \Cassini spacecraft during 16 orbits
between July 2004 and September 2005. The mean standoff distance for this interval
was found to be $\sim\unit[25]{\RS}$. We thus adopt this value for our present work
as an appropriate magnetopause radius for a nominal magnetodisc model representing average
solar wind conditions at Saturn. 
The presence of the magnetopause boundary requires a corresponding contribution to the magnetic
potential $\alpha$ from the currents flowing on that boundary. \citet{caudal} represented
this magnetopause potential at Jupiter as the Euler function corresponding to a globally uniform, southward-directed 
field $\vec{B}_S$,
referred to as the `shielding field'. \citet{caudal} chose the magnitude of $\vec{B}_S$ by requiring
that the magnetic flux due to the shielding field,  integrated over the entire equatorial plane,
be equal to a prescribed fraction $\xi$ of the total
magnetic flux exterior to the boundary due to the planetary plus disc sources. The addition of the
shielding potential to the solution for $\alpha$ at each iteration thus `compresses' the flux tubes of the outer
magnetosphere inwards from their `boundary-free' configuration. 

\begin{figure}
\centering
\includegraphics[width=0.99\figwidth]{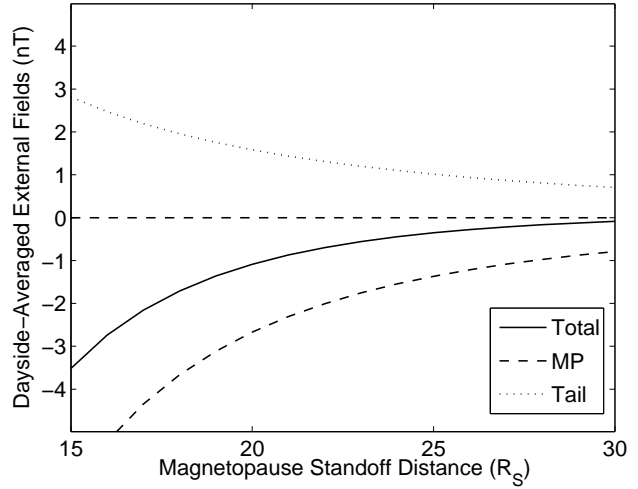}
\caption{
Dayside-averaged equatorial values of the shielding magnetic fields associated with Saturn's magnetotail current sheet 
(`Tail') and magnetopause currents (`MP'). The field value shown is the $Z$ component, i.e.\
positive northward.
The global fields used for the averaging were calculated using the formulae
of \citet{alexeev2006} for the configuration where the planetary dipole is orthogonal to the upstream solar wind.
} 
\label{fig:shield_field}
\end{figure}

For the magnetopause contribution in our axisymmetric models, we adopted a similar
approach to \citet{caudal}; however, we determined our value of the uniform field $\vec{B}_S$ by performing dayside
equatorial averages of the empirical field models described by \citet{alexeev2005,alexeev2006}, which represent contributions
from both the magnetopause and magnetotail current sheets at Saturn. These two contributions  
are oppositely-directed (magnetopause field southward, magnetotail field northward).
We computed our shielding field as a function of $\RMP$, using the following 
parameters to represent approximate conditions at Saturn, as required in the expressions of \citet{alexeev2006}:
(i) planetary dipole orthogonal to the solar wind flow direction,
(ii) radial distance $\RT$ of the inner edge of the tail sheet equal to \unit[0.7]{\RMP},
(iii) magnitude of field in the tail lobe given by
$B_L = \Phi_L \RMP^{-2} / 
\left[\frac{\pi}{2}(1 + 2 \RT / \RMP) \right]$, with open magnetic flux
$\Phi_L = \unit[40]{GWb}$. 
The resulting magnetopause contributions, before global averaging, showed variation by a factor
$\sim2\mbox{--}3$ between noon local time (strongest field) and the dawn / dusk meridian. The magnetotail contribution
showed similar relative variability but with the strongest fields situated at dawn / dusk. Thus in the full representation
there are local times where the two contributions add to zero. The uniform (dayside-averaged) 
shielding field used in our model
is shown as a function of $\RMP$ in \Fig{\ref{fig:shield_field}}. We show the contributions
to the total shielding field from the magnetopause and tail currents. It is evident that for Saturn the magnetopause
currents dominate the shielding field for the more compressed magnetosphere. For the more expanded
configuration, the presence of tail currents significantly decreases the shielding field magnitude below 
its predicted values from magnetopause currents alone. 
 
\begin{figure}
\centering
\includegraphics[width=0.99\figwidth]{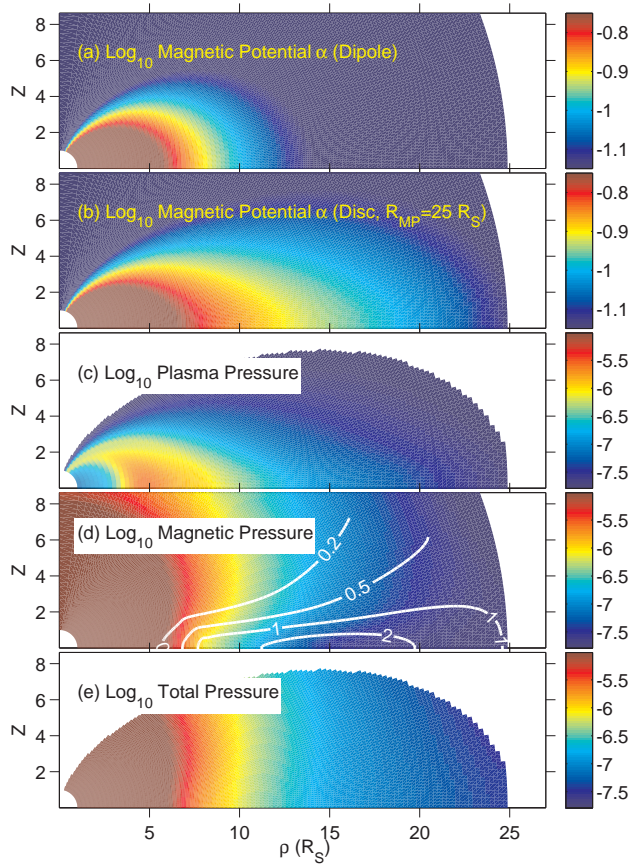}
\caption{
(a) Logarithm of magnetic potential $\alpha$ for vacuum dipole configuration, shown on 
a colour scale as a function of cylindrical co-ordinates $\rho$ and $Z$;
(b) Logarithm of magnetic potential for an average magnetodisc field model for Saturn,
with magnetopause radius \unit[25]{\RS} and hot plasma index
$K_{h} = \unit[2\cdot10^{6}]{Pa\,m\,T^{-1}}$ (see text);
(c) Distribution of plasma pressure within the model from (b);
(d) Distribution of magnetic pressure within the model from (b), along
with labelled contours of constant plasma $\beta$ (thick white lines);
(e) Total (plasma plus magnetic) pressure on a colour scale. The approximately
vertical pressure levels near the equatorial plasma disc are a consequence of
force balance perpendicular to the magnetic field (see text).
} 
\label{fig:disc_av_struc}
\end{figure}

Several output parameters associated with our average model ($\RMP=\unit[25]{\RS}$)
for the Kronian magnetodisc are depicted
on a colour scale in the panels of \Fig{\ref{fig:disc_av_struc}}. \Fig{\ref{fig:disc_av_struc}}(a) shows contours of
constant magnetic potential $\alpha$, equivalent to magnetic field lines, 
for the vacuum dipole used to
represent Saturn's internal field in our model. We may compare
this geometry with the average magnetodisc model in \Fig{\ref{fig:disc_av_struc}}(b) 
which corresponds to magnetopause
radius $\RMP=\unit[25]{\RS}$. The radial stretching of field lines  
compared to the dipole model becomes particularly pronounced
beyond $\sim\unit[8]{\RS}$.
For example, the magnetic flux contained between the equatorial distances $\unit[6\mbox{--}10]{\RS}$ in the dipole
field becomes spread out over a larger interval  $\unit[8\mbox{--}18]{\RS}$ in the full 
magnetodisc solution. We shall
compare the equatorial field profiles for these models later in this section.

We now consider \Fig{\ref{fig:disc_av_struc}}(c), which shows the distribution of total plasma pressure
in the $(\rho,Z)$ plane. The scale length $\lsc$ for the model ranges between $\unit[{1\mbox{--}5}]{\RS}$
through the magnetosphere, monotonically
increasing with $\rho$. The pressure contours which attain separations from the equatorial plane
significantly larger than these scales are 
primarily due to the hot plasma pressure, which we have assumed to be uniformly distributed along field lines.
One can also see the influence of the equatorial confinement of the cold population, by comparing
individual contours with the field line shapes: the pressure contours tend to be more oblate. 
\Fig{\ref{fig:disc_av_struc}}(d) shows the magnetic pressure distribution, along
with contours of plasma $\beta$, which clearly show the influence of the equatorial confinement of the cold
population for $\beta$ of the order unity or larger. The contours of magnetic pressure turn inwards towards
the planet as they approach the equator. This is a consequence of force balance perpendicular to the radially-stretched
field lines just outside the equatorial plasma disc (e.g.\ \citet{mgkdjs2005}). The main forces acting in this direction (which is approximately
perpendicular to the equator) are the plasma and magnetic pressure gradients. To maintain balance
as the disc is approached, the corresponding increase in plasma pressure must be balanced by decrease in
magnetic pressure; hence the behaviour of the magnetic pressure contours. We thus expect total plasma plus
magnetic pressure to be constant along the vertical direction near the disc. \Fig{\ref{fig:disc_av_struc}}(e)
shows contours of this total pressure, and confirms that they follow directions nearly perpendicular to the equator.

\begin{figure}
\centering
\includegraphics[width=0.99\figwidth]{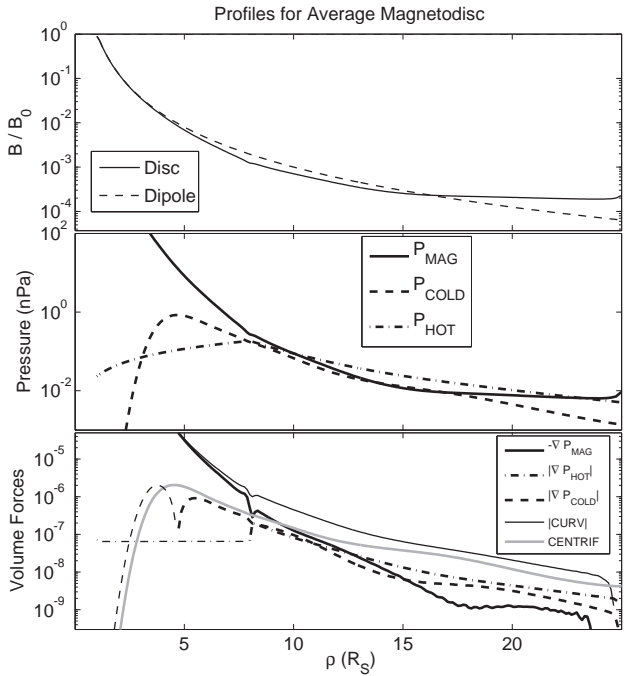}
\caption{
Upper panel: Equatorial profiles of magnetic field strength for the planetary dipole alone
and for the full magnetodisc solution ($\RMP=\unit[25]{\RS}$).
Middle panel: equatorial model profiles of magnetic and plasma pressures
(hot and cold). Bottom panel: normalised volume forces in the equatorial plane of the
model, labelled according to line style. 
We show the absolute value of force.
Thick lines indicate positive (outward) radial force, while thin lines
show regions where the force is inwardly-directed (negative).
} 
\label{fig:discavprofs}
\end{figure}

We shall continue our present investigation of average plasmadisc structure at Saturn by considering 
the model's equatorial properties of magnetic field
and force balance in \Fig{\ref{fig:discavprofs}}.
The upper panel compares the equatorial profiles of magnetic field strength
associated with the planetary internal dipole, and with our full magnetodisc solution for average
magnetopause size. As for the simple zeroth-order disc models (\S\ref{sec:zdisc}), the presence
of the plasmadisc produces a total field profile somewhat weaker than the parent dipole for the
regions closest to the planet, and stronger than dipole field beyond a characteristic transition distance.
The middle panel of the figure shows equatorial profiles of magnetic pressure, cold plasma 
pressure and hot plasma pressure. We note that the magnetic pressure exceeds that of the plasma for
distances smaller than $\sim\unit[10]{\RS}$. The hot pressure is the dominant source for
distances around $\sim\unit[15]{\RS}$.
The bottom panel of \Fig{\ref{fig:discavprofs}} shows the equatorial
profile of the absolute value of the various volume forces. 
We emphasise here that we have used 
line thickness to indicate regions where radial forces are directed 
outward (thicker lines) or inward (thin lines).
Over most of the model magnetosphere the
curvature force is the principal, inward-directed (i.e.\ negative radial) force. The sum
of all the radial forces in the equatorial plane has a magnitude less than 0.2 percent of the local curvature force; 
this fraction thus provides some measure of the degree of accuracy with which the model can simulate perfect force balance.

The bottom panel of \Fig{\ref{fig:discavprofs}} also indicates which forces dominate the balance and determine
disc structure in different regions of the equatorial magnetosphere. Throughout the magnetosphere, the magnetic curvature 
force is the strongest inward-directed force. For distances $\rho\gtrsim\unit[15]{\RS}$, centrifugal force is higher
than plasma pressure gradients by factors up to five, and is therefore the second most important term in the disc's
stress balance. Closer to the planet, for $\rho\sim\unit[6\mbox{--}12]{\RS}$, centrifugal force and plasma
pressure gradients are comparable in magnitude, and the disc's field structure is determined by both
sources of radial stress in approximately equal measure. These calculations
are in broad agreement with the conclusions of \citet{arridge2007} who used
current sheet crossings to show that centrifugal and pressure gradient
forces were approximately equal in magnitude at \unit[20]{\RS} whereas the model shows the centrifugal forces slightly larger at about twice that of the pressure gradient forces.

Our average Kronian disc model contains a 
hot plasma pressure distribution which is indicative of a `mildly disturbed' ring current 
(see \Fig{\ref{fig:input_hot_press}}). We therefore would expect hot plasma pressure
to play a more dominant role in magnetospheric force balance under conditions 
of so-called `disturbed' ring current, as shown by the \Cassini observations \citep{sergis2007}.
We defer a detailed investigation of this aspect to a future study, and concentrate here on modelling conditions
characteristic of the mean level of observed hot pressure.

\Fig{\ref{fig:discavprofs}} shows a small `kink' in the magnetic force profiles around
\unit[8]{\RS}; this is due to the sharp linear decrease we have assumed for
characterising the product of hot plasma pressure and unit flux tube volume 
(\S\ref{sec:hotplasmapressure}). The termination at this distance
of the curve representing the
outward-directed force due to hot plasma pressure
confirms a sharp change in the sign of the hot pressure gradient;
this feature in turn corresponds to the
rapid decline with decreasing distance of the modelled hot plasma density. 
The kink feature is thus somewhat artificial,
but does not affect the validity of the global features of our modelled force profiles.

We now consider the inner magnetospheric region ($\rho\lesssim\unit[6]{\RS}$) depicted
in \Fig{\ref{fig:discavprofs}}. Inside this distance, the cold plasma population density rapidly
decreases (as also shown by the behaviour of the centrifugal force, which is proportional to cold plasma pressure).
This magnetospheric region is then characterised by a relative absence of plasma
and a magnetic field dominated by the planetary dipole.

We conclude this section on the average magnetodisc structure at Saturn by investigating the
relationship between the previously-considered forces which act to create the magnetodisc
geometry and the magnetospheric currents which flow in response to the presence of those
forces. On a microscopic scale, we expect the main azimuthal currents to arise from drift
motions associated with: (i) finite plasma pressure (gradient and curvature drifts),
(ii) centrifugal force associated with plasma rotation (inertial current)
(e.g.\ \citet{bunce2007}). The macroscopic
formalism of the model allows an alternative identification
of these currents from force balance considerations, as follows.
For our Saturn model, 
the relevant scaling factor for $J_{\phi}$ is listed in \Table{\ref{scales_appendix_table}}. 
In order to separate the contribution of a particular force
to the current density, we simply substitute its corresponding contribution
to the source function \Eq{\ref{eq:zdisc_g}} for the function
$g$ used in \Eq{\ref{eq:jphifromg}}.

\begin{figure}
\centering
\includegraphics[width=0.99\figwidth]{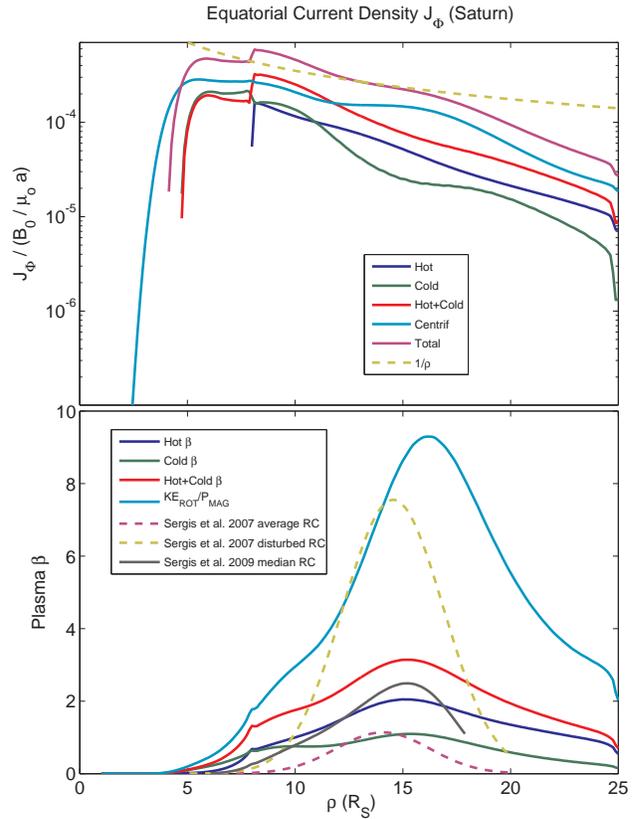}
\caption{
Upper panel: Equatorial profiles of positive azimuthal current density
taken from the Saturn magnetodisc model with $\RMP=
\unit[25]{\RS}$. The profiles are plotted on a logarithmic scale
and are colour-coded according to the
force with which they correspond in the plasmadisc's dynamical balance
(hot / cold plasma pressure, centrifugal force). A function proportional to 
$1/\rho$ is also shown in order to indicate the form assumed for the
current density in the CAN annular disc model \citet{connerney1981};
Lower panel: Equatorial profiles of plasma $\beta$ 
taken from the Saturn magnetodisc model with $\RMP=
\unit[25]{\RS}$. Profiles are colour-coded according to the physical
origin of the energy density used to compute the $\beta$ ratio
(hot / cold plasma pressure, rotational kinetic energy). The dashed
and solid grey curves show the plasma $\beta$ profiles fitted to
observations of the hot plasma pressure at Saturn
obtained by \citet{sergis2007,sergis2009}.
} 
\label{fig:disc_av_eqjphi}
\end{figure}

Following this method, we calculated the various contributions to
azimuthal current in the equatorial plane of the average magnetodisc model. 
Profiles of positive $J_{\phi}$ (in the direction of planetary rotation)
are shown on a logarithmic scale in the upper panel
of \Fig{\ref{fig:disc_av_eqjphi}}. It is clear that the force
which is associated with the dominant
contribution to the magnetospheric current depends on radial distance.
For example, we note that there is a broad local maximum in the centrifugal
inertial current centred at $\sim\unit[16]{\RS}$. This feature
corresponds to a similar local maximum in plasma angular velocity
according to the model profile from \Fig{\ref{fig:plasma_omega}}.
The lower panel of \Fig{\ref{fig:disc_av_eqjphi}} shows equatorial
profiles of plasma $\beta$ along with an equivalent $\beta$ for the rotating disc
plasma, computed as the ratio of the rotational kinetic energy density to
the magnetic pressure. This rotational plasma $\beta$ also peaks near
$\sim\unit[16]{\RS}$, thus indicating that the dominant term to the
plasma source function (and therefore azimuthal current) in this region is 
the centrifugal force term. Rotational $\beta$ then decreases for
$\rho > 16$ due to the decline in cold plasma density. The corresponding
effect on the
current density profiles is a smaller ratio 
in the outer magnetosphere of the centrifugal to plasma current density.

In the region $\rho\sim\unit[8\mbox{--}12]{\RS}$, we see from
\Fig{\ref{fig:disc_av_eqjphi}} that the current due to total
plasma pressure gradient slightly exceeds the centrifugal current. 
The hot plasma current is an important factor here; the observed
strong variability in hot plasma pressure at Saturn (\S\ref{sec:hotplasmapressure},
\citet{sergis2007,krimigis2007}) implies that differing levels of
ring current activity may plausibly increase the radial extent of
this region where plasma pressure dominates magnetospheric current, or even
lead to its {\em disappearance\/}. Inside $\unit[8]{\RS}$ the hot plasma density sharply decreases
and the corresponding decrease in the associated current profile produces an inner
region where centrifugal current is once more the major contribution. We shall defer
a detailed investigation of the influence of hot plasma index $K_h$ (\S\ref{sec:hotplasmapressure})
on magnetospheric current profiles to a future study. For present purposes, we note that 
the calculations indicate it is expected to play a significant role in determining the extent
of the region where hot plasma pressure is the major source of the azimuthal current density.

Alongside the modelled plasma $\beta$ 
in the lower panel of \Fig{\ref{fig:disc_av_eqjphi}}, we also show observed values of
hot plasma $\beta$ presented by \citet{sergis2007,sergis2009} (grey solid and dashed curves).
The dashed grey curves indicate fits to hot plasma observations by \citet{sergis2007},
which show the variation in hot plasma $\beta$ between average and
disturbed ring current states (see also \S\ref{sec:hotplasmapressure}).
The solid grey curve was determined from a more recent fit to the median equatorial values 
of
hot plasma $\beta$ at Saturn determined by \citet{sergis2009} 
(computed over $\sim\unit[0.1]{\RS}$ intervals), who included the significant contribution
($\sim50$ percent) to hot pressure due to \chem{O^+} ions. If we compare this curve with
the median profile from the
earlier study (light grey dashed curve), we see that this inclusion has significantly increased the
hot plasma $\beta$ which would characterise an average state of the ring current. Given this result,
and the large intrinsic variability in observed hot plasma pressures, we believe that our simplified
model distribution of hot plasma $\beta$ (red curve) is in reasonable agreement with the expected
global behaviour of this parameter. For the region $\rho\gtrsim\unit[20]{\RS}$ our modelled
hot plasma $\beta$ is in excess of the declining values of \citet{sergis2007,sergis2009}; however, the
distant magnetospheric observations by \citet{krimigis2007} (see \Fig{\ref{fig:input_hot_press}})
do show hot plasma $\beta$ which are consistent with our choice for $K_h$. Improved future determinations
of plasma moments will no doubt enable us to further refine our plasma parametrisation , but for the present study we 
shall remain with the description given in \S\ref{sec:hotplasmapressure}.

Another important feature of the plasma $\beta$ profiles in 
\Fig{\ref{fig:disc_av_eqjphi}} is the relatively uniform ratio of $\sim2\mbox{--}3$ in the outer magnetosphere
($\rho\gtrsim\unit[15]{\RS}$) between hot and cold plasma $\beta$. Our calculations also
show that in this region the length scale $\lsc$ for the cold disc plasma (\Eq{\ref{eq:ldefn}})
monotonically increases with distance between $\sim\unit[3\mbox{--}5]{\RS}$.
If we use these values in \Eq{\ref{eq:rho_trans}} for the transition distance in a homogeneous
plasma disc between pressure- and centrifugally-dominated regions, we obtain  
$\rho_T\sim\unit[12\mbox{--}22]{\RS}$. This range of transition distances is consistently smaller
than the model magnetopause radius. The actual transition distance for the model appears
to be situated at $\sim\unit[12]{\RS}$, beyond which distance the centrifugal current
persistently exceeds total plasma pressure current. The homogeneous disc
predictions for $\rho_T$ thus suggest that the actual value
of this distance is expected to lie within Saturn's magnetosphere, and the full magnetodisc model
confirms that this is indeed the case. The formula in \Eq{\ref{eq:rho_trans}} therefore provides
a reasonable means of estimating the order of magnitude of $\rho_T$ from observed and / or theoretical
properties of the magnetospheric plasma.

\begin{figure}
\centering
\includegraphics[width=0.99\figwidth]{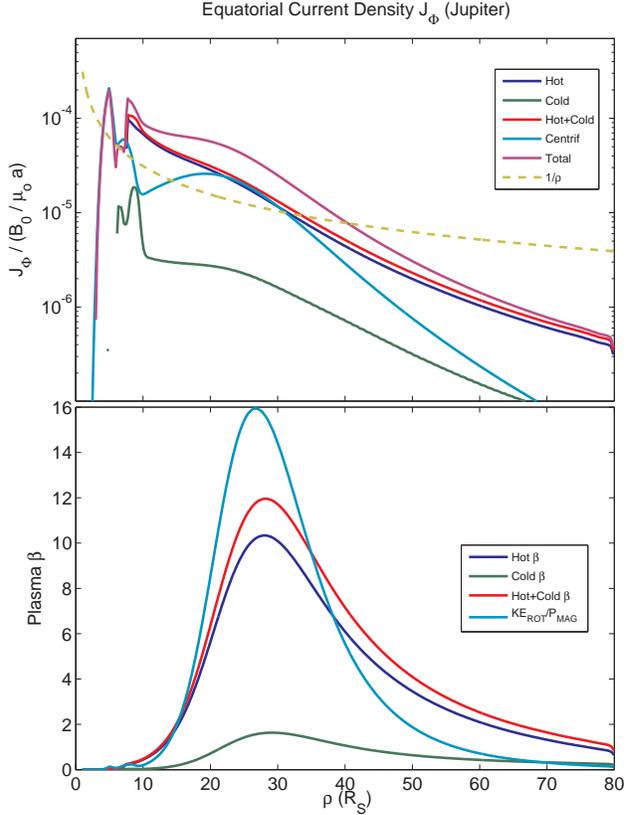}
\caption{
Upper panel: Equatorial profiles of positive azimuthal current density
taken from the Jupiter magnetodisc model with $\RMP=
\unit[80]{\RJ}$ (reproduction of the calculation by \citet{caudal}). 
The profiles are plotted on a logarithmic scale, and are colour-coded according to the
force with which they correspond in the plasmadisc's dynamical balance
(hot / cold plasma pressure, centrifugal force). A function proportional to 
$1/\rho$ is also shown in order to indicate the form assumed for the
current density in the CAN annular disc model \citet{connerney1981};
Lower panel: Equatorial profiles of plasma $\beta$  
taken from the Jupiter magnetodisc model with $\RMP=
\unit[80]{\RJ}$. Profiles are colour-coded according to the physical
origin of the energy density used to compute the $\beta$ ratio
(hot / cold plasma pressure, rotational kinetic energy). 
} 
\label{fig:disc_jup_av_eqjphi}
\end{figure}

We now make a comparative investigation of general magnetodisc structure by comparing the profiles
in \Fig{\ref{fig:disc_av_eqjphi}} for Saturn's plasmadisc with those shown in
\Fig{\ref{fig:disc_jup_av_eqjphi}} for Jupiter. The results in 
\Fig{\ref{fig:disc_jup_av_eqjphi}}
reproduce the model calculation by \citet{caudal} for a Jovian magnetodisc
with magnetopause radius $\RMP=\unit[80]{\RJ}$. The most striking difference
between the Jupiter model and
Saturn model is the clear dominance of the Jovian outer
magnetosphere's equatorial current density by 
hot plasma pressure. The hot plasma current is the major contribution
to total $J_{\phi}$ for distances beyond $\sim\unit[40]{\RJ}$.
We also note a much stronger contrast between hot and cold plasma $\beta$ for Jupiter compared to
Saturn. While the ratio $\beta_h / \beta_c$ is an order of magnitude or more beyond
$\sim\unit[40]{\RJ}$ in the Jovian model, the same quantity is $\lesssim2$ in the Kronian
calculation. As a result, the current profiles due to hot and cold plasma pressure gradients show generally
comparable values at Saturn, while at Jupiter the cold plasma current is an order of magnitude or more
weaker compared to that of the hot plasma.

These results indicate that the much more expanded magnetosphere of Jupiter develops an outer
region beyond 
$\sim\unit[40]{\RJ}$,
where the cold plasma's angular velocity and density decline 
at a rate sufficiently rapid to produce a plasma whose main energy content
arises from the thermal motions of the hot particle population. Near
$\sim\unit[27]{\RJ}$ in the Jovian model, the rotational plasma $\beta$
exceeds the hot plasma $\beta$ and the centrifugal current becomes
comparable with the hot plasma
current. This is qualitatively similar to the corresponding behaviour near $\sim\unit[16]{\RS}$
in the Kronian model. If we repeat the exercise of computing the transition distance 
for the values of plasma $\beta$ and scale length from the Jovian model,
we obtain values of $\rho_T$ in excess of $\RMP$
(the values for length scale are $\lsc = \unit[5\mbox{--}40]{\RJ}$, increasing with distance). 
This indicates that the centrifugal current at Jupiter should never exceed
the hot plasma current in the outer magnetosphere, according to the simple homogeneous disc model.
The full magnetodisc model we have presented for Jupiter confirms this prediction, showing a hot plasma-dominated
magnetospheric current beyond \unit[40]{\RJ}.

We now consider
the relative magnitudes of the magnetospheric current at Jupiter and
Saturn predicted by the models. Both \Fig{\ref{fig:disc_av_eqjphi}}
and \Fig{\ref{fig:disc_jup_av_eqjphi}} show normalised current densities,
expressed using scale factors of 
$\unit[280]{nA\,m^{-2}}$ (Saturn)
and $\unit[4800]{nA\,m^{-2}}$ (Jupiter) (see \Table{\ref{scales_appendix_table}}). 
Although the absolute value of the scale
current at Jupiter is much higher because of that planet's stronger internal field, we note 
something interesting when we
compare the normalised current densities at both planets within the same distance range of
$<25$ planetary radii: the values of normalised $J_{\phi}$ at Saturn over 5--16 planetary radii 
exceed those at Jupiter by factors of $\sim5$. Since the distance scale is similar
for both models, we conclude that this feature is an indication that Saturn's ring current produces
a stronger {\em relative\/} perturbation to the planet's internal dipole within this distance range.
Interestingly, \citet{vasyliunas2008} arrived at a similar conclusion by considering the 
plasma outflows near the orbital
distances of Io and Enceladus ($\sim6$ and $\sim4$ planetary radii, respectively)
and demonstrating that these flows would be expected to
produce a stronger relative distortion of the planetary dipole for Saturn.

Our model calculations also show a spatial profile of total $J_{\phi}$ in the
outer magnetosphere, for both Jupiter and Saturn, which falls off more steeply
with radial distance $\rho$ than the $1/\rho$ dependence used by the 
CAN current disc model. This is an important point of comparison, as it indicates
that an outer plasmadisc structure obeying radial stress balance has a 
characteristic spatial gradient in current density which is significantly different to that
usually assumed in ring current modelling studies. Despite this difference,
however, both the Caudalian and CAN disc models
are suitable for reproducing the larger-scale observed structures in the magnetodisc
field, as we shall see in the following sections. The main advantage of the Caudalian
disc is that it also provides realistic spatial profiles of current and radial force
arising from {\em self-consistent\/} global distributions of plasma.

\subsection{Response of Magnetodisc to Solar Wind Pressure}
\label{sec:disc_response}
\begin{figure}
\centering
\includegraphics[width=0.99\figwidth]{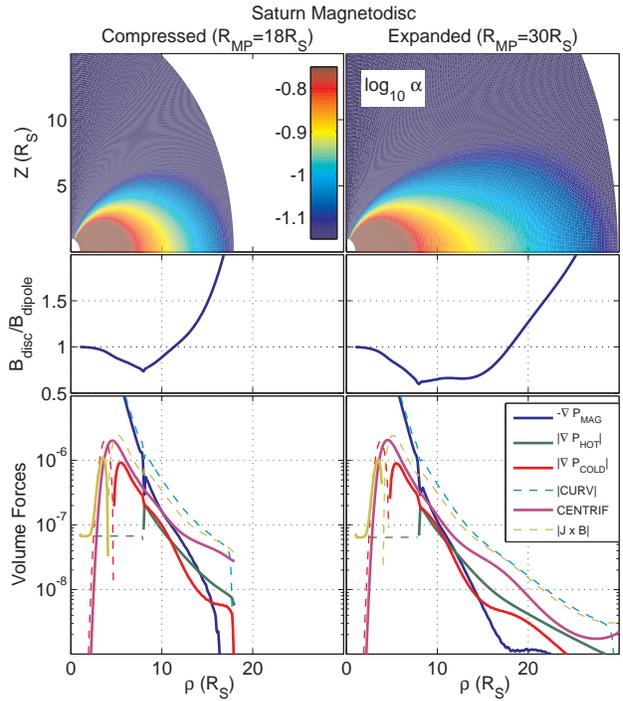}
\caption{
The left and right columns of plots correspond respectively to Saturn disc models calculated 
for compressed  ($\RMP=\unit[18]{\RS}$) and
expanded  ($\RMP=\unit[25]{\RS}$) configurations.
Top panels: The logarithm of magnetic potential $\alpha$ is plotted on a colour scale for the labelled
configurations.
Middle panels: The equatorial ratio of total to dipole magnetic field strength is plotted for both 
magnetodisc configurations. The increased field strength of the compressed magnetodisc is apparent.
Bottom panels: Equatorial profiles of the absolute value of
normalised volume forces for the compressed and expanded 
models, labelled according to line colour. Line style is used to indicate the
direction of the radial forces,  with solid lines indicating outward force and
dashed lines indicating inward force.
} 
\label{fig:discdiffprofs}
\end{figure}

In this section, we parametrise the effect of solar wind dynamic pressure by varying the magnetopause
radius $\RMP$ in our model calculations. In \Fig{\ref{fig:discdiffprofs}},
we present model outputs calculated for two 
configurations. The first corresponds to strongly
compressed conditions for the Kronian magnetosphere with $\RMP=\unit[18]{\RS}$,
and the second is for a value $\RMP=\unit[30]{\RS}$ which is 
typical of the most expanded
magnetospheric structures observed in the \Cassini era
\citep{achilleos2008,arridge2006}.
We emphasise that the plasma parameters of temperature, angular
velocity, flux tube content and
hot plasma index are identical in the two models. The final solution for the 
magnetic field within each model will change the mapping between these
last two parameters and local quantities, such as number density and pressure, according to 
the frozen-in condition.

We commence our comparison of the compressed and expanded magnetodisc structures
by considering the top panels of \Fig{\ref{fig:discdiffprofs}} which show
contours of constant magnetic potential $\alpha$, equivalent to field line shapes.
The region of strongly-radial field near the equatorial plane, as seen in the average model 
(\S\ref{sec:avg_mdisc}), is also present in the expanded
disc, particularly in the range $\rho\sim\unit[15\mbox{--}20]{\RS}$. 
The compressed magnetodisc, on the other hand, displays field line shapes which are
far less radially `stretched' and which more closely resemble the geometry of a pure dipole
(see \Fig{\ref{fig:disc_av_struc}}). 
A similar result was found by \citet{bunce2008} who modelled the ring
current for various magnetospheric configurations as revealed by \Cassini magnetometer data
from a selection of orbits.
The colour scale of the upper 
panels in  \Fig{\ref{fig:discdiffprofs}} indicates that both
compressed and expanded models have similar levels of magnetic flux threading their
entire equatorial planes; we therefore expect higher field strengths to be present in the
compressed disc. The middle panels confirm that this is the case.
Equatorial profiles of total magnetic field strength relative to that of the planetary dipole 
are shown as a function of $\rho$. Beyond $\sim\unit[5]{\RS}$, the compressed
disc model has a persistently stronger magnetic field than the expanded one. Around
$\sim\unit[15]{\RS}$, for example, the compressed field has reached a magnitude twice as large
as the expanded configuration.

This behaviour of the field strength and geometry under strongly-compressed conditions has
important consequences for the ensuing magnetic forces which operate within the plasmadisc.
In the bottom panels of \Fig{\ref{fig:discdiffprofs}}, we plot equatorial profiles of
the volume forces due to plasma pressure gradients, magnetic pressure gradient, magnetic 
curvature and centrifugal force. 
The plots show that magnetic curvature is the principal, radially-inward force for both
disc configurations. Closer inspections of the two curvature force profiles reveals a remarkable feature;
the compressed model shows a stronger curvature force beyond $\sim\unit[8]{\RS}$,
whose ratio with respect to the expanded disc attains a maximum of
$\sim1.4$ at $\rho \sim \unit[15\mbox{--}17]{\RS}$. The compressed
model is able to maintain a stronger curvature force via higher magnetic
field strength, despite the increased radius of curvature of the local field line.
We also show plots of the total magnetic force $\vec{J}\times\vec{B}$
for both models (sum of curvature force and magnetic pressure gradient). 
A comparison of the two sets of curves reveals that, beyond
$\sim\unit[8]{\RS}$, the magnetic pressure gradient in the compressed disc is
larger relative to the curvature force than in the expanded case.
This behaviour is qualitatively consistent with the study by 
\citet{arridge2008} mentioned in \S\ref{sec:intro}, 
which showed that the dayside magnetospheric field at Saturn
only becomes significantly `disc-like' under conditions of low solar wind dynamic pressure 
($\RMP>\unit[23]{\RS}$). This aspect is also in accordance with the conclusions
of \citet{bunce2008}.

Within the range of radial distances $1 < \rho < \unit[18]{\RS}$ covered by the compressed model's equatorial plane,
there are also significant differences in magnetic pressure gradient and centrifugal force
with respect to the expanded model. Firstly, the magnetic pressure within this
distance range falls off with distance more gradually in the compressed disc.
For both configurations, power-law fits to the magnetic pressure, 
$P_\mathrm{MAG}\propto\rho^{-2\chi}$,
were obtained for the interval   $10 < \rho < \unit[15]{\RS}$.
The resulting indices were $\chi=2.80\pm0.14$ (compressed) and
$\chi=3.27\pm0.10$ (expanded), revealing that the expanded model field
falls off slightly more rapidly than a pure dipole $\chi=3$ in this region.
However, a similar fit to the apparently more uniform part of the 
expanded field strength profile in the more distant
magnetosphere $20 < \rho < \unit[25]{\RS}$ yielded $\chi=1.12\pm0.08$.
These results indicate that the compressed Kronian outer magnetosphere is likely to be
characterised by field strength gradient similar to that of a dipole, while a more
expanded configuration may be expected to exhibit a field with a more gradual
decline, associated with values of the index $\chi$ in the range 1--3. This predicted
behaviour of the magnetospheric field suggests that observational
studies of the relationship between magnetopause standoff distance and solar wind pressure
may benefit from the assumption of a field strength index $\chi$ which varies with
$\RMP$, rather than the usually assumed fixed value (e.g.\
\citet{achilleos2008,arridge2006,slavin1985}).

If we now turn our attention to the centrifugal force profiles in 
\Fig{\ref{fig:discdiffprofs}}, a detailed inspection reveals
that the compressed model exhibits a centrifugal force consistently
stronger than that of the expanded disc for $\sim8 < \rho < \unit[18]{\RS}$,
with the ratio of the two increasing monotonically to a value of $\sim2$.
This is a consequence of the higher cold plasma densities in the compressed
model (at a given $\rho$, the ratio of centrifugal force between the two
configurations is equivalent to the ratio of cold plasma density).
In the region $\sim8 < \rho < \unit[15]{\RS}$, 
the plasma pressure gradients in the two models differ by less than
\unit[10]{per\,cent}.
The increased centrifugal force of the compressed disc is thus balanced by
an increased magnetic force (difference between magnetic curvature inward and
magnetic pressure gradient outward). Interestingly, the region 
$\sim15 < \rho < \unit[18]{\RS}$ near the compressed magnetopause
is characterised by a change in sign of the magnetic pressure gradient, which is
required to maintain balance due to the curvature force decreasing more rapidly
than the centrifugal force.

The region $\sim15 < \rho < \unit[20]{\RS}$ for the expanded magnetodisc
has force balance mainly determined by magnetic curvature and centrifugal effects
as shown by \Fig{\ref{fig:discdiffprofs}}. The same region, however,
has a broad local maximum in centrifugal force 
nearly coincident with a minimum in magnetic pressure gradient.
These features arise because of the corresponding local maximum of plasma 
angular velocity in the same region (\Fig{\ref{fig:plasma_omega}}), and
the field geometry producing a relatively uniform region of field strength.
In the more distant magnetosphere near $\sim\unit[23\mbox{--}27]{\RS}$ 
the hot plasma
pressure gradient becomes equal in importance to centrifugal force in maintaining
force balance due to the declining density and angular velocity of the cold disc plasma.
As for the case of the compressed disc, the magnetic pressure near the magnetopause
in the expanded model begins to increase with distance in order to maintain force
balance near the boundary.

\subsection{Comparison of Model to Magnetic Field Observations}
\label{sec:model_data}
A comprehensive comparison of our magnetodisc model for Saturn with the
vast field and plasma datasets from the \Cassini spacecraft is beyond the scope of this
paper.
For the sake of a preliminary assessment of how well the model may be applied
to spacecraft observations, we shall present in this section a comparison between
the magnetodisc model field and magnetometer observations from the \Cassini spacecraft
from two quite different orbits. The first is the \rev{3} orbit lying entirely within Saturn's rotational
equator, from the early part of the mission (February~2005) and the second is the highly-inclined 
\rev{40} orbit from
March~2007 which sampled the entire vertical structure of the disc.

\begin{figure}
\centering
\includegraphics[width=0.99\figwidth]{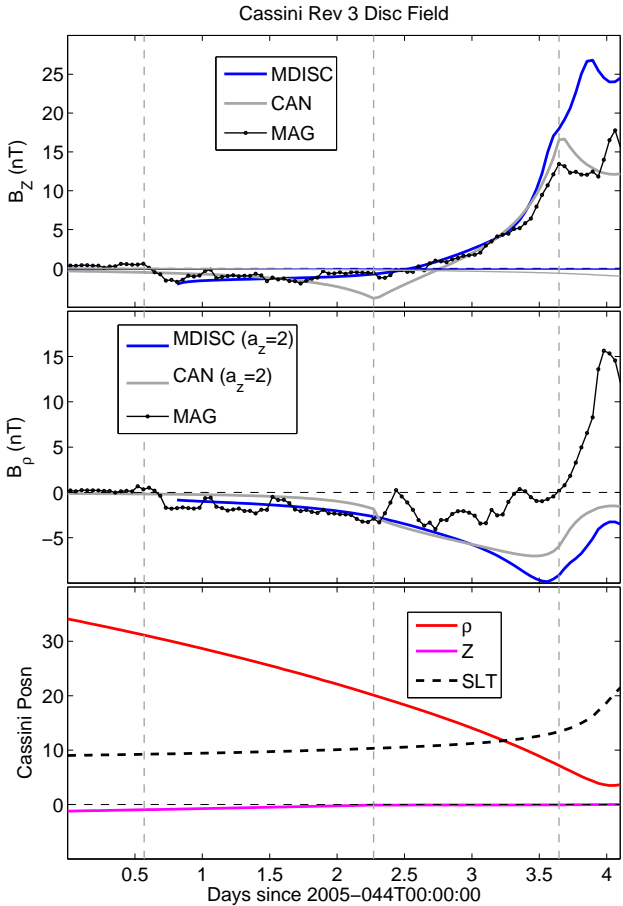}
\caption{
The top two panels show a comparison between modified \Cassini magnetometer
data from the \rev{3} orbit (`MAG') and
models which employ the Connerney (`CAN') and Caudalian (`MDISC') discs, with parameters as described in the text. 
The thin coloured curves in the top panel represent the small contributions
in the models from magnetopause and tail currents. Vertical ($\BZ$) and
radial ($\Brho$) field components are shown as a function of time. The data are hourly averages and have had the internal
field model described by \citet{dougherty2005} subtracted.  The middle panel uses a magnetic equator for the models which is
displaced by \unit[2]{\RS} north of the planet's rotational equator 
(without such a displacement, the models' predicted
values for $\Brho$ would be identically zero for this equatorial orbit). The bottom panel shows the spacecraft position as a function of time 
(cylindrical radial ($\rho$) and vertical ($Z$) distance, \SLT). The vertical dashed lines indicate, from left to right,
the positions of the last inbound magnetopause crossing, the outer edge of
the CAN model disc (\unit[20]{\RS}) and 
the inner edge of the same model
(\unit[7]{\RS}).
The CAN disc parameters were taken from \citet{bunce2007}, who fit a 
non-displaced model to observations of $\BZ$.
}
\label{fig:mdisc_vs_data_rev03}
\end{figure}

\Fig{\ref{fig:mdisc_vs_data_rev03}} depicts the information relevant for our comparison 
based on \Cassini \rev{3}, covering a period of approximately four days
in February~2005. The time axis is labelled in days since the beginning of Day of Year~44, or February~13.
The bottom panel shows the position of \Cassini as a function of time using the colour-coded $\rho$ and $Z$ 
cylindrical coordinates as well as the \SLT in decimal hours. We see that this inbound
pass of the orbit sampled the magnetosphere at radial distances $\rho \sim \unit[4\mbox{--}31]{\RS}$,
the largest distance in this range corresponding to the indicated magnetopause crossing.
The orbital segment for which $\rho\gtrsim\unit[8]{\RS}$ was situated at near-noon
local times between $\sim\unit[10\mbox{ and }14]{h}$ (\SLT). The entire orbit was
also situated
within or very close to the equatorial plane $Z=0$. This region of space is thus appropriate for
analysis with our model, which is representative of dayside conditions at Saturn and which uses a
simplified formulation for the field due to
magnetopause currents, based on an empirical model of the dayside equatorial field due to this
source (\S\ref{sec:avg_mdisc}).

%
%

The upper panels of \Fig{\ref{fig:mdisc_vs_data_rev03}} show the $\Brho$ and $\BZ$
components of the magnetic field (black curve) observed by \Cassini during the relevant time interval, in \unit{nT} units,
from which we have subtracted the components of the internal field model described by
\citet{dougherty2005}. The plotted data thus represents the magnetic field due to the external sources of 
the current disc and magnetospheric boundaries.
To analyse the $\BZ$ observations, we chose two models. The first
is the non-homogeneous part of our magnetodisc
model (i.e.\ the total field model minus the planetary dipole term) with hot plasma index 
$K_h = \unit[2\cdot10^6]{Pa\,m\,T^{-1}}$ (representing approximately
average ring current activity at Saturn) and appropriate magnetopause radius 
$\RMP=\unit[30]{\RS}$. The value for $\RMP$
based on the magnetopause crossing location and the magnetopause model of \citet{arridge2006}
is \unit[28]{\RS} --- the use of either value did not significantly change the results. 
The $\BZ$ values shown by the blue curve were
obtained by linear interpolation of the field values from our 2D model grid onto the spacecraft trajectory.
We also show, using grey curves, the predictions from the CAN model used by \citet{bunce2007}
to analyse these data. The parameters for this model are the azimuthal scale current per unit radial length
$I_0$, the inner and outer edges of the annular model disc ($a$ and $b$) and the disc half width in the
$Z$ direction ($D$). The parameter values we chose were the following, as determined by \citet{bunce2007}:
$\mu_0I_0= \unit[53.3]{nT}$, $a = \unit[7]{\RS}$, $b = \unit[20]{\RS}$,
$D = \unit[2.5]{\RS}$. 

The thin blue and grey curves show the corresponding contributions
from the magnetopause shielding field to the different disc models. 
On the scale of the plot, the uniform
shielding field of our model (see \Fig{\ref{fig:shield_field}}) has a barely discernible magnitude of
\unit[0.09]{nT}. As for our Caudalian model, the shielding field for the CAN model was 
assumed to lie entirely in the $Z$ direction, but was computed using the following formula
from \citet{bunce2007}:
\begin{align}
\BZ^\mathrm{MP} = \frac{B_1(X-X_2) + B_2(X_1-X)}{X_1-X_2}.   \label{eq:bunce_bzmp}
\end{align}
This expression describes a shielding field which changes linearly with $X$, the spatial coordinate
associated with the axis which lies along the intersection of the equatorial plane and the noon-midnight
meridian ($X$ positive towards the Sun). The 
parameter values $B_1=\unit[0]{nT}$ and $B_2=\unit[-1.11]{nT}$ were
chosen to fit the observed $\BZ$ values at the position of the magnetopause ($X_1=\unit[23.36]{\RS}$) 
and the nightside location with the minimum value of $X$ ($X_2=\unit[-6.17]{\RS}$).
It is evident that the shielding field for both models makes only a very minor contribution to the total
predicted field except, for the CAN model, in the region adjacent to the magnetopause.

Structure at a variety of timescales is evident in the observations. The global nature of the disc models
implies that they are suitable for analysing the largest scales, of the order half a day in time or
a few planetary radii in $\rho$. If we firstly consider the $\BZ$ field, both models reasonably reproduce
the overall trend seen in the observations. Near the location of the outer edge of the CAN model
(vertical line at $\sim\unit[2.3]{days}$), we see
that this model predicts a relatively sharp minimum in $\BZ$ due to the truncated nature of its current
disc. The Caudalian disc with its extended current sheet makes a smoother transition in $\BZ$ through
this region in better agreement with the observations. On the other hand, near the location of the inner
edge of the CAN model (vertical line at $\sim\unit[3.7]{days}$) the local peak in $\BZ$ displayed by this
model fits the data more closely than the Caudalian disc, which rises to values more than twice that
of the data within this inner region. This suggests a need for more accurate plasma inputs in the 
region $\rho\lesssim\unit[5]{\RS}$ of our model, as discussed in \S\ref{sec:plasmacompn}
and \S\ref{sec:coldplasma}. The local peak in the $\BZ$ data near $\sim4$ days is most likely a 
signature of the `camshaft' field at Saturn. This is a quasi-periodic modulation seen in the magnetic field whose
physical origin remains to be unambiguously identified, but appears to be linked with field-aligned,
azimuthally-modulated magnetospheric
currents flowing between the ionosphere and plasmadisc (e.g.\ \citet{djsmgk2007,provan2009}). 
We shall return to this aspect 
when we consider the high-latitude observations.

Both disc models 
fail to agree with the $\Brho$ data in \Fig{\ref{fig:mdisc_vs_data_rev03}}.
This is because they have a north-south hemispherical symmetry, which by definition requires
$\Brho=0$ within the equatorial plane. The fact that \Cassini observes a significantly
non-zero $\Brho$ in \rev{3} and many other equatorial orbits has been suggested
to be the result of a non-planar plasmadisc structure; in particular, the bowl-shaped current sheet
model explored by \citet{arridgewarp2008} provides an explanation for these observations.
Such a sheet morphology would be expected to arise in a magnetosphere where the planetary dipole is significantly
non-orthogonal with respect to the upstream solar wind flow direction, as
was the case during \rev{3}, where the
angle between these two directions was $\sim\unit[70]{^\circ}$ (the northern magnetic pole being tilted away from the
Sun). As a first, albeit crude, approximation to the ensuing field geometry,
the model $\Brho$ values shown 
are those corresponding to a model plasmadisc which has been displaced by a distance of 
\unit[3]{\RS}
north of Saturn's rotational equator. 
This displacement is consistent with the predictions of the
current sheet model by \citet{arridgewarp2008} for a distance $\rho = \unit[25]{\RS}$
and a subsolar latitude $\unit[-23]{^\circ}$, appropriate for the time of the \rev{3} orbit.
The $\Brho$ field of our model in the region $\rho = \unit[20\mbox{--}30]{\RS}$ changed by
$\lesssim\unit[0.5]{nT}$ when we changed the displacement by \unit[1]{\RS}. We shall
address the variability of the current sheet displacement with $\rho$ in a future study. For present
purposes, we use \unit[2]{\RS} as a representative displacement for this outer magnetospheric region.

In the region between the magnetopause boundary and the neighbourhood of
the outer edge of the CAN disc, both models and data are in reasonable agreement, confirming that a displaced
planar disc is a useful representation of the local effects of the more realistic bowl-like shape. As we proceed closer
to the planet along the spacecraft orbit towards the inner CAN disc edge, the observed $\Brho$ decreases in
magnitude, consistent with the magnetic equator of the plasmadisc becoming aligned with the rotational equator;
the displaced model values, unsurprisingly, do not fit the data in this region. We see a local peak in the observed
$\Brho$ near \unit[4]{\RS} corresponding to the similar feature in $\BZ$.
We noted that this peak in $\BZ$ changed by $\sim\unit[5]{nT}$ if we used a
different internal field model for subtraction \citep{burton2009}. Thus we cannot provide a definitive
explanation for this feature without further examination of the internal field models used in the near-planet region.

\begin{figure}
\centering
\includegraphics[width=0.99\figwidth]{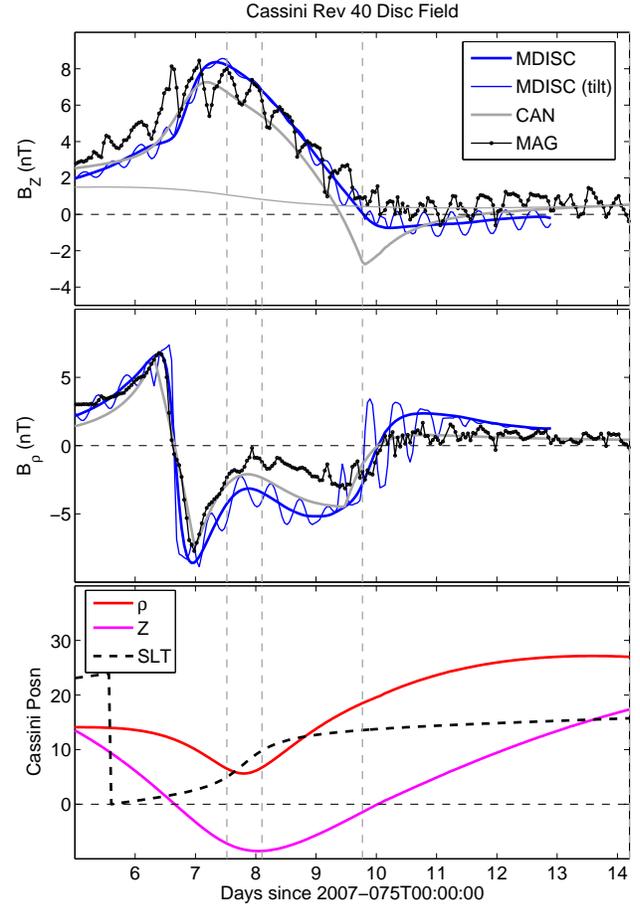}
\caption{
Plots analogous to \Fig{\ref{fig:mdisc_vs_data_rev03}}
for the \Cassini \rev{40} orbit.
Predictions for a rotating disc whose axis is tilted at $10^{\circ}$  to that
of the planet's rotation axis are also shown (`MDISC (tilt)'). 
The vertical dashed lines indicate the positions of the
inner (\unit[6.6]{\RS}) and outer (\unit[18.6]{\RS}) radii of the CAN disc, 
whose parameters have been chosen to best fit 
the $\Brho$ data (see text). The time axis is truncated on the right at the first 
outbound magnetopause crossing.
} 
\label{fig:mdisc_vs_data_rev40}
\end{figure}

In the panels of \Fig{\ref{fig:mdisc_vs_data_rev40}}, we plot magnetic field components and
spacecraft position as a function of time using the same scheme and conventions 
as \Fig{\ref{fig:mdisc_vs_data_rev03}}. The time interval in question covers about nine days
from the beginning of March~21, 2007 which correspond to the closest approach to Saturn and outbound
segment of the \rev{40} orbit. The $Z$ co-ordinate trace in spacecraft position shows that during this
time \Cassini probed regions up to \unit[15]{\RS} from the equatorial plane, and the spacecraft latitude
reached magnitudes of $\sim60^\circ$. In addition, at the time intervals near 6.5 and 10~days, the spacecraft
traversed the full extent in $Z$ through the current sheet, a structure with typical vertical length scales of a few
\unit{\RS}. From the discussion in \S\ref{sec:avg_mdisc}, typical length scales along the $\rho$
co-ordinate for the cold disc plasma are $\sim\unit[1\mbox{--}5]{\RS}$, which therefore 
provide an upper bound for the length scale along $Z$. 
\rev{40} thus provides a very different view of the magnetosphere compared to the
equatorial pass of \rev{3} and hence a good means of further testing the suitability of the disc models
for magnetic analyses.
 
%

We shall consider firstly the $\BZ$ data and model predictions in the top panel of 
\Fig{\ref{fig:mdisc_vs_data_rev40}}. As for \rev{3}, structure in the magnetic
field on a variety of time scales is seen; in particular, the quasi-periodic ($\sim\unit[10.75]{hr}$)
camshaft signal in $\BZ$ is clearly evident with typical amplitudes of the order
\unit[1]{nT}. We shall return to this feature presently after discussing the larger-scale
features in the field profile. The Caudalian and CAN disc models used for comparison
purposes are shown as thick blue and grey curves. 
We chose the following parameters for the CAN disc, obtained from a least-squares fit
to the combination of both observed field components as displayed in the figure:
$\mu_0I_0= \unit[40]{nT}$, $a = \unit[6.6]{\RS}$, $b = \unit[18.6]{\RS}$,
$D = \unit[3.2]{\RS}$. 
The thin grey curve
shows the shielding field profile used in the CAN model, computed using the following parameters
for \Eq{\ref{eq:bunce_bzmp}}:
$B_1=\unit[0.5]{nT}$, $B_2=\unit[1.5]{nT}$,
$X_1=\unit[15]{\RS}$, $X_2=\unit[-14]{\RS}$. The thin blue curve shows a modified
version of the Caudalian disc, which we describe in more detail later in this section.
We note that the nightside subset of these data was modelled by \citet{kellett2009} using the CAN
formulation. These authors used a substantially thinner current disc ($D = \unit[0.4]{\RS}$) and
a correspondingly more intense current parameter ($\mu_0I_0= \unit[338]{nT}$) to optimally fit the nightside field.

The location of the outbound magnetopause
crossing at $(\rho,Z) = (26.8, 17.3)\unit{\RS}$ corresponds to a subsolar
standoff distance $\RMP=\unit[24]{\RS}$, using the axisymmetric magnetopause
model of \citet{arridge2006}. However, we found that the Kronian disc model with a somewhat
larger magnetopause radius
$\RMP=\unit[30]{\RS}$ gave significantly better agreement with the observations,
and it is the field profiles for this more expanded model which we have displayed.
Since the magnetopause crossing was at a relatively high altitude $Z$ above the equator, this finding may
indicate that the magnetopause of Saturn exhibits polar flattening, although evidence for
this requires further studies of similar mid- to high-latitude boundary crossings. 

For the interval spanning closest approach until the outbound magnetopause crossing, 
both the empirical CAN model and the physical Caudalian model
reproduce the large-scale trend in $\BZ$ from the magnetometry. The Caudalian disc predicts a mean
field about \unit[1]{nT} weaker than that observed in the time interval after 10~days. This feature
suggests that a positive shielding field, similar to that employed for the CAN model, may be a more realistic
choice for this pass than the uniform negative value \unit[-0.09]{nT} used in our model (see \Fig{\ref{fig:shield_field}}). 
The camshaft signal in $\BZ$ is observed throughout this orbit. This field source, when added to the planetary
dipole, has been suggested to be
equivalent to that of a tilted, rotating dipole in the outer magnetosphere $\rho\gtrsim\unit[15]{\RS}$
\citep{djsmgk2007}. In this picture, we would expect the magnetic equator of the outer Kronian plasmadisc
to also be tilted relative to the rotational equator. As a preliminary exploration of this concept, we have plotted
in \Fig{\ref{fig:mdisc_vs_data_rev40}} a thin blue curve showing the field profiles associated with a
tilted, rotating plasmadisc. We computed these profiles by simply taking the original magnetodisc model
and transforming it to a co-ordinate system where the model's axis of cylindrical symmetry is tilted at an angle
of $10^{\circ}$ with respect to the planet's rotation axis (the latter now being defined as the $Z$ direction
in accordance with the data). The orientation of the model symmetry axis was also allowed to vary with time
such that the azimuthal angle of its projection onto the rotational equator corresponded to a regular rotation with
a period of \unit[10.75]{h}. We note that the observed camshaft signal does not have a fixed period, but one which may drift
in value by the order of a minute over time scales of the order of a year, as revealed by its radio signature,
e.g.\ \citet{kurth2007,kurth2008}. 
We include the tilted disc calculations here simply to emphasise that such a model
\emph{cannot} be consistent with the observations in their entirety.
Although the amplitude of tilted disc $\BZ$ fluctuations
match the data on the outbound pass reasonably well for $\rho\gtrsim\unit[10]{\RS}$, they 
rapidly diminish inside this region. By contrast, the observations show persistent field fluctuations
throughout the orbit. These quasi-periodic fluctuations, i.e.\ the `camshaft signal',
the phase relations between the different components, and their origin, have been the subject
of much research \citep{espinosa2000,cowley2006,djsmgk2007,provan2009}. These studies
also highlight the difference between phase relations of the camshaft field components and 
those of a simple rotating, tilted disc.


\section{Summary and Discussion}
\label{sec:summary}
We have introduced a new model for Saturn's magnetodisc, based on an original formalism by \citet{caudal}.
The model formalism is based on the magnetostatic
solution for an Euler potential consistent with global balance between plasma pressure gradient, centrifugal
force and magnetic force ($\vec{J} \times \vec{B}$) in a cylindrically symmetric system. Such an approach
has the advantage of being able to predict a self-consistent system of plasma properties, magnetospheric azimuthal
currents and magnetic field. The equatorial boundary condition for the model was provided by observations from the
\Cassini spacecraft of hot and cold plasma pressure, and cold plasma density and temperature (\S\ref{sec:modelinputs}
and subsections). In order
to formulate a model with realistic global behaviour, we adopted relatively simple functional forms for these
physical parameters. In this context, the unit flux tube volume concept was applied, following \citet{caudal},
in order to compute global distributions of plasma whose density and pressure would respond appropriately to different
magnetospheric radii, according to the behaviour expected of a `frozen-in' plasma. The empirical fits to hot plasma pressure,
cold plasma composition and cold plasma temperature 
by \citet{sergis2007,sergis2009,wilson2008} were employed in order to achieve a reasonable representation for the
model equator of average magnetospheric conditions at Saturn. We also used a polynomial fit to the data by
\citet{kane2008,wilson2008} for plasma angular velocity.

Before presenting the outputs from the full model, we considered a simple toy model emphasising the
largest angular scales of the magnetic potential for a homogeneous disc, characterised by constant plasma $\beta$, constant
plasma scale length $\lsc$ and full corotation with the parent planet. This simple model was used to show the influence on
magnetic field geometry to be expected when a rotating plasmadisc is added to a planetary dipole. In particular, hot 
plasma pressure generally inflates outer magnetospheric flux tubes to greater radial distances while the centrifugal
confinement of the rotating cold plasma towards the equator gives rise to inflated, relatively oblate field lines.
This zeroth-order behaviour was consistent with our full model for Saturn's magnetodisc under average internal
(ring current activity) and external (solar wind) conditions.
Our consideration of this baseline
model, for which magnetopause radius $\RMP=\unit[25]{\RS}$
confirmed the radial `stretching' of the unperturbed dipolar field lines as a result of the currents flowing 
mainly in the equatorial plasmadisc. The corresponding equatorial field strength in the model
shows a region where it falls below the unperturbed dipole value for distances
$\sim\unit[5\mbox{--}15]{\RS}$ and exceeds the dipole value beyond this range, also displaying a 
comparatively more gradual decrease with
radial distance in the outer magnetosphere. All of these features are general characteristics which arise
from adding the solenoid-like magnetic field of the disc (ring) current alone to the planetary dipole.

Examination of the equatorial radial forces in the average Kronian disc model revealed that, for
distances beyond $\sim\unit[15]{\RS}$, the principal forces determining disc structure are the
magnetic curvature and centrifugal forces. This characteristic distance is consistent with the simple
formula for the `transition distance' $\rho_T$ between pressure- and centrifugal-dominated
structure which arose from the zeroth-order disc treatment.
This formula reveals the conditions under which
$\rho_T$ is most likely to exceed the magnetopause radius, and consequently the plasmadisc can never have a force
balance dominated by centrifugal force. The relevant conditions are:
(i) hot plasma $\beta$ is very high compared to the cold plasma (i.e.\ thermal energy is large
compared to rotational kinetic energy), (ii) plasma angular
velocity is adequately low, or (iii) for a given temperature of cold plasma, its density is small (such
that the quantity $\lsc^2/\beta_c$ becomes very large.

Consideration of the equatorial, azimuthal current density $J_{\phi}$ in the average Kronian disc model
revealed that centrifugal inertial current was the primary contribution for distances beyond
$\sim\unit[13]{\RS}$. For the region $\sim\unit[8\mbox{--}12]{\RS}$, hot plasma and centrifugal
current were predicted to be comparable. However, a further exploration of more active ring current
states (future study) is likely to show that this interval in distance will expand, and the hot plasma current
intensify, as the hot plasma index
$K_h$ 
is increased beyond values appropriate for average conditions at Saturn.

A comparison of the azimuthal current profiles between the average Saturn disc model and a reproduction
of the Jovian magnetodisc by \citet{caudal} (for which $\RMP=\unit[80]{\RJ}$)
was also revealing.
In particular, the calculations confirmed that the increased $\beta_h/\beta_c$ ratio at Jupiter endows
this planet's magnetosphere with equatorial azimuthal current dominantly due to hot plasma pressure 
beyond a distance $\sim\unit[30]{\RJ}$. Within $\sim\unit[20\mbox{--}30]{\RJ}$, the centrifugal and
hot plasma currents are of similar magnitude. 

The normalised (dimensionless) quantities adopted in
our model enabled us to make an important comparison between the strength of the azimuthal
current density at Saturn and Jupiter. The values of normalised equatorial $J_{\phi}$ at Saturn, according to
our calculations, are expected to exceed those at Jupiter by factors of $\sim$ 5 within the
distance range  $\sim\unit[5\mbox{--}16]{}$ planetary radii. The implication of this result is that, while the
absolute strength of the Kronian currents is far weaker than their Jovian counterparts, the {\em relative\/}
perturbation to Saturn's internal field in this distance range due to the disc current would exceed that at Jupiter.

In \S\ref{sec:disc_response}
we examined the response of the Saturn disc model to conditions of
compressed ($\RMP=\unit[18]{\RS}$)
and expanded ($\RMP=\unit[30]{\RS}$)
magnetospheric configuration. Both models generally showed centrifugally-dominated force balance beyond
$\sim\unit[15]{\RS}$, although the expanded disc shows comparable centrifugal force 
and hot plasma pressure gradient near $\sim\unit[25]{\RS}$ due to the decline in cold plasma angular velocity.
Interestingly, the compressed disc is able to maintain a similar or stronger curvature force than the expanded one, despite having
nearly dipole-shaped field lines: this property is a consequence of the higher field strengths attained in the compressed
magnetospheric state. Around
$\sim\unit[15]{\RS}$, for example, the compressed model's equatorial
field already reaches a magnitude twice as large
as the expanded configuration. 
Consideration of the gradients in magnetic pressure in the compressed and expanded
disc models indicated that the index $\chi=-\frac{\rho}{B}\deriv{B}{\rho}$,
which characterises the relative change in field strength $B$ per relative
change in radial distance $\rho$, is likely to vary as a function of magnetopause standoff 
distance. This dependence recommends
the corresponding use of a variable $\chi$ in future observational studies of the response of
Saturn's magnetopause boundary to changing solar wind conditions.

Finally, in \S\ref{sec:model_data}, we presented model calculations for a Kronian disc model with
$\RMP=\unit[30]{\RS}$ and average hot plasma index. We compared
our model predictions for vertical ($\BZ$) and radial ($\Brho$) field components with
magnetometer data from two of the orbits of the \Cassini spacecraft's prime mission.
We also presented model calculations for appropriate CAN
annular disc models \citep{connerney1981} as part of this
comparison. In general, both the Caudalian and CAN disc model were able to 
account for the general large-scale trends seen in the data-derived magnetic field due to
the magnetodisc current alone. However, certain discrepancies between the models and
the observations point to a need to use non-planar disc geometries in more detailed studies.
The first of these discrepancies is the non-zero radial field observed by \Cassini
during the \rev{3} orbit considered herein, which cannot be explained by a disc field
with north-south hemispheric symmetry.
An observational study by \citet{arridgewarp2008} for many equatorial orbits revealed that this
is a repeatable signature, and is most likely associated with a bowl-shaped current sheet.

The second important discrepancy between our model calculations and the magnetometry
is the presence of observed quasi-periodic fluctuations in the field, known as the camshaft
signal. While our rotating, tilted disc model was able to qualitatively reproduce similar
field fluctuations in the outer magnetosphere ($\rho\gtrsim\unit[15\mbox{--}20]{\RS}$), it
was clearly not capable of explaining the observed behaviour of the camshaft signal for regions
closer to the planet. A general advantage of the Caudalian disc model in the context of data
interpretation is that its more realistic plasma distribution yields smoother predicted changes
in field orientation for spacecraft passes through the current sheet. The CAN model predicts
sharp peaks in field components during such transitions due to its assumption of
an annular geometry with definitive boundaries for the current-carrying region; it also
shows similar abrupt changes in field for the regions near the assumed inner and outer edges
where the current region is truncated. The discussion in \S\ref{sec:avg_mdisc} revealed that the force balance
used to derive the Caudalian disc structure also results in a fall-off in magnetospheric
current density more rapid than the $1/\rho$ law assumed in the CAN model.

The Caudalian magnetodisc for Saturn represents a useful first model for pursuing studies of
the plasmadisc structure, azimuthal current and magnetospheric field, along the lines that
we have presented in this paper. While these initial studies have revealed some interesting
features of disc structure and currents at Saturn, particularly when compared to the Jovian
system, they also highlight some important future directions for work involving this model,
such as the following.
\begin{enumerate} 
\item 
Improved determinations of plasma moments should be incorporated into the structure of
the model, in order to provide more accurate depictions of the global plasma conditions.
\item
Investigation of the influence of hot plasma pressure on magnetodisc structure. Our
initial study has revealed that it plays a potentially important role in determining the general
structure of the magnetodisc field and the extent of the magnetospheric region where the
electric current density $J_{\phi}$ is dominantly determined by energetic particle
motions, rather than the inertial current associated with centrifugal force acting on the cold 
population.
\item
An extension of our preliminary study of solar wind influence on disc structure to additional
magnetopause radii and global characterisations of internal plasma energy and content. The
self-consistent response of plasma angular velocity to magnetospheric compression could also
play a potentially important role here.
\item
Further analyses of \Cassini field and plasma data. The spacecraft has thus far completed
more than 100 orbits of Saturn. Such a vast dataset will require much time to exploit. A suitable use
for our model with regard to the field and particle data would be a modelling study of selected orbits
using input plasma moments acquired during those orbits, rather than a `global approximation' to these
conditions. The model outputs would thus reflect conditions most appropriate for the orbits in question.
Such calculations would be of use, for example, to teams who aim to derive magnetospheric particle fluxes and
current densities directly from {\em in situ\/} measurements.

\end{enumerate}

\section*{Acknowledgements}
The authors acknowledge the continued support and collaboration of the \Cassini magnetometer (MAG), plasma spectrometer
(CAPS) and magnetospheric imager teams (MIMI). CSA was supported for part of this work by an STFC rolling grant at
MSSL. PG was supported for part of this work by an STFC rolling grant at UCL Physics and Astronomy. NA wishes to thank
Nick Sergis, Cesar Bertucci and Stephanie Kellett for useful discussions. The authors thank the 
referee S.~W.~H.\ Cowley for very comprehensive and helpful comments and suggestions.

\bibliographystyle{mn}
\bibliography{References}

\appendix

\section{Solutions for the Magnetodisc Potential}
\label{sec:mdiscsoln}
This Appendix describes the derivation of the solution for the magnetic
potential of an axisymmetric plasma distribution given in articles by
\citet{caudal} and \citet{lackner1970}.
The potential in question is denoted by $\alpha$ and is actually one of two Euler potentials from which the magnetic field \vec{B} may be derived
\begin{align}
\vec{B} = \vec{\nabla} \alpha \times \vec{\nabla} \beta,
\end{align}
where $\alpha$ is a function of radial distance $r$ and cosine of colatitude $\mu = \cos \theta$. 
The function $\beta$ depends only on azimuthal angle $\phi$: $\beta = a \phi$ with $a$ being the planetary radius. 
Note that many pairs of Euler potentials can be associated with a particular magnetic field, however this particular choice separates the azimuthal 
and meridional dependencies. In effect, these equations tell us that an individual magnetic field line can be thought of 
as the line of intersection of a surface of constant $\alpha$ (which will resemble a `doughnut-shaped' shell) and a plane of constant $\beta$ 
(which is simply the meridional plane with azimuth $\phi$).

Let us now consider the differential equation for the meridional Euler potential $\alpha$ given by \citet{caudal}
( we shall use the dimensionless system of co-ordinates described in this paper): 
\begin{align}
\pdsec{\alpha}{r} + \frac{1-\mu^2}{r^2} \pdsec{\alpha}{\mu} = -g(r,\mu,\alpha).
\label{eq:caudal_de_merid}
\end{align}

In \Eq{\ref{eq:caudal_de_merid}}, the function $g$ represents a source term describing a distribution of 
external plasma and currents which must 
be specified \emph{a priori}. Note that $g$ requires knowledge of $\alpha$
as well --- the function we are trying to solve for. 
\citet{caudal} and \citet{lackner1970}  solve this problem through an iterative approach. 
One starts with an initial `guess' $\alpha_0$ for the functional form of $\alpha$. 
$\alpha_0$ is then used to evaluate $g$, and then
\Eq{\ref{eq:caudal_de_merid}} is solved to give an
updated solution $\alpha_1$.  $\alpha_1$ is then used in the next iteration to re-evaluate $g$ and 
to update the solution again. The process is repeated until convergence: in practice, one usually stops when the
maximum relative difference between successive iterations falls below some user-defined tolerance.

A reasonable first guess for $\alpha$ is the planetary dipole potential
\begin{align}
\alphadip (r, \mu) =  \frac{1-\mu^2}{r}.  \label{eq:dip_potl}
\end{align}
$\alphadip$ is a homogeneous solution of \Eq{\ref{eq:caudal_de_merid}} (i.e.\ a solution for the case where the source term is identically zero). 
However, it is not the only homogeneous solution. 
Homogeneous solutions are a 
good starting point for finding \emph{particular} solutions (i.e.\ when the source term is non-zero).
We may obtain the general form of the homogeneous solution by using the property of separability 
i.e.\ $\alpha (r,\mu) = \alpha_r (r) \alpha_\theta (\mu)$ is the product of two single-variable functions
as stated previously. Substituting this into \Eq{\ref{eq:caudal_de_merid}} we can
show that
\begin{align}
\frac{r^2}{\alpha_r} \dsec{\alpha_r}{r} + \frac{1-\mu^2}{\alpha_\theta}
\dsec{\alpha_\theta}{\mu} = 0.
\label{eq:caudal_de_homo}
\end{align}

Now if we fix the value of $r$, we would expect the left-hand term in this equation (a function of $r$ only) to be a constant.
However the equation tells us that this constant is independent of whatever value of $\mu$ we use to evaluate the right-hand term. 
It follows that the right-hand term, regardless of the value of $\mu$, must be a constant. A similar argument, keeping $\mu$ fixed, 
reveals that the left-hand term, for all values of $r$, must also be equal to a constant. If the constant takes on special values, derivable 
from an integer $n \geq 0$, we may write
\begin{align}
\frac{r^2}{\alpha_r} \dsec{\alpha_r}{r} & =  n(n+3)+2,
\label{eq:radial_de_homo}\\
- \frac{1-\mu^2}{\alpha_\theta} \dsec{\alpha_\theta}{\mu} & = n(n+3)+2.
\label{eq:angular_de_homo}
\end{align}

It is easy to show that the radial part $\alpha_r$ has a solution of the form
\begin{align}
\alpha_r = C_l r^l,
\end{align}
where $C_l$ is a constant, and integer $l$ must satisfy $l(l-1) = n(n+3)+2$. 
Solving this quadratic, we see that $l$ can take on the value $n+2$ or $-(n+1)$. $l=n+2$ 
corresponds to a positive power of $r$ and a potential which monotonically
increases with distance from the planet --- this is not physical. 
Therefore, we choose $l = -(n+1)$ for the radial part of the function. 

To solve for the angular function $\alpha_\theta$ we require knowledge of the Jacobi polynomials. The particular strand of these polynomials 
which are of use to us here are denoted 
\jacpoly{n}{\mu} ($n$ is an integer $\geq 0$ so we can associate a polynomial with each choice of $n$ in equations
\ref{eq:radial_de_homo} and \ref{eq:angular_de_homo}). The useful property of the polynomials \jacpoly{n}{\mu} is that 
the functions $(1-\mu^2) \jacpoly{n}{\mu}$ are actually solutions of 
\Eq{\ref{eq:angular_de_homo}}. 
For a given choice of $n$ our homogeneous solution would thus be
\begin{align}
\alpha_H(r,\mu) = \alpha_r \alpha_\theta = C_n r^{-(n+1)} (1-\mu^2)
\jacpoly{n}{\mu}.
\end{align}

Since \Eq{\ref{eq:caudal_de_homo}} is linear in $(\alpha_r \alpha_\theta)$, it follows that any linear combination of solutions of the above 
form is also a homogeneous solution. Without loss of generality, the complete homogeneous solution is thus
\begin{align}
\alpha_H(r,\mu) = (1-\mu^2) \sum_{n=0}^{\infty} C_n r^{-(n+1)}
\jacpoly{n}{\mu}.
\label{eq:soln_homo}
\end{align}

We have now found solutions for the \emph{homogeneous} (source-free) version of Caudal's equation. But how 
do we use these to obtain a solution for the full differential
\Eq{\ref{eq:caudal_de}} which contains the source function $g$? 
We try a general solution obtained by multiplying each term in the series of the homogeneous solution by a purely radial function $f_n(r)$. 
This trial function thus takes the form
\begin{align}
\alpha(r,\mu) = (1-\mu^2) \sum_{n=0}^{\infty} f_n(r) r^{-(n+1)}
\jacpoly{n}{\mu},
\label{eq:fullalpha}
\end{align}
where we have absorbed the constant $C_n$ into the definition of $f_n(r)$. 
Now if we use this trial solution in the left-hand side of
\Eq{\ref{eq:caudal_de}}, we obtain
\begin{align}
(1-\mu^2) \sum_{n=0}^{\infty} 
\left(\dsec{f_n}{r} - \frac{2(n+1)}{r} \deriv{f_n}{r} \right)
r^{-(n+1)} \jacpoly{n}{\mu} \nonumber\\
= g(r,\mu,\alpha_{i-1}).
\label{eq:alphaexpansion}
\end{align}

Here we have introduced the symbol $i$ to emphasise that solving this equation is 
part of an iterative process where the 
solution $\alpha_i$ is obtained 
from the previous one $\alpha_{i-1}$. Although the left-hand side of our equation retains 
the form of a series summation in the Jacobi polynomials, we can't progress much further without 
addressing the right-hand side.  This is where the Jacobi polynomials again prove useful. 
They are an orthogonal, complete set of functions, which means that \emph{any} function of $\mu$ can be 
expressed as a series expansion using Jacobi polynomials. Applying this to our function $g$ for an arbitrary value of 
radial distance $r$, we can decompose the angular dependence of $g$ into a sum over the polynomials as follows
\begin{align}
g(r,\mu,\alpha_{i-1}) = (1-\mu^2) \sum_{n=0}^{\infty} g_n(r)
\jacpoly{n}{\mu}, \label{eq:gexpansion}
\end{align}
with the expansion coefficients defined by the orthogonality condition
\begin{align}
g_n(r) & = \frac{1}{h_n} \int_{-1}^{1} g(r,\mu) \jacpoly{n}{\mu} d\mu, \\
h_n & = \int_{-1}^{1} (1-\mu^2) (\jacpoly{n}{\mu})^2 d\mu.
\label{eq:orthog_condn}
\end{align}

We can now make use of the orthogonality of the polynomials to equate the
$n$-th terms of \Eqs{\ref{eq:alphaexpansion}--\ref{eq:gexpansion}}. This gives
\begin{align}
\left(\dsec{f_n}{r} - \frac{2(n+1)}{r} \deriv{f_n}{r} \right) r^{-(n+1)}   = -g_n(r),
\end{align}   
or equivalently (multiplying both sides by $r^{-(n+1)}$)
\begin{align}
r^{-2(n+1)}\dsec{f_n}{r}-2(n+1)r^{-(2n+3)}\deriv{f_n}{r}
= - r^{-(n+1)} g_n(r).
\end{align}   

We see that the left-hand side can be expressed as the derivative of a product
\begin{align}
\deriv{}{r} \left(r^{-2(n+1)}f_n^{\prime}\right) = - r^{-(n+1)} g_n(r).
\label{eq:fn_de}
\end{align}
The left-hand side of this equation is readily integrable. But we see that the general solution for the $f_n(r)$ functions will 
involve integrals of the source function $g$. What this means in practice is that we have to numerically integrate some kind of 
empirical or other function which is a fit to observed plasma distributions. \citet{caudal}'s work shows that the source function 
includes quantities such as plasma pressure, plasma temperature (assumed isotropic) and mean ion mass. 
We now finalise the integration towards a final solution.
We start with \Eq{\ref{eq:fn_de}} and rename the dummy variable for radial distance to $u$
\begin{align}
\deriv{}{u} \left(u^{-2(n+1)} f_n^{\prime} \right) = - u^{-(n+1)} g_n(u) .
\label{eq:fnu_de}
\end{align}

We now integrate both sides over the range $r_c$ to $r$. $r_c$ is an inner boundary, similar to the planetary radius, 
which encloses the region where the field is purely a dipole field i.e.\ purely due to the planet's internal source. 
We adopt the boundary condition that $f_n$ = 0 at $u = r_c$ (i.e.\ the contributions to the potential from the plasma source disappear at the inner
boundary)
and $f_n^{\prime} = f_c^{\prime}$ at $u=r_c$ (there is a `jump' in the potential gradient at the inner boundary $u = r_c$
supported by currents flowing on that surface). Performing this integration between $u = r_c$ and $u = r$ gives us
\begin{align}
f_n^{\prime} = r^{2(n+1)} \left( f_c^{\prime} r_c^{-2(n+1)} -  G(r) \right),
\label{eq:fn_prime}
\end{align}
where $G(r)$ denotes the function
\begin{align}
G(r) &= \int_{r_c}^{r} u^{-(n+1)} g_n(u)  du, \\
\deriv{G}{r} &= r^{-(n+1)} g_n(r) .
\end{align}

If we integrate \Eq{\ref{eq:fnu_de}} between the limits $u = r_c$ and $u = \infty$ we obtain the useful identity
\begin{align}
f_c^{\prime} r_c^{-2(n+1)} = G(\infty).
\label{eq:fc_prime}
\end{align}

We can now integrate \Eq{\ref{eq:fn_prime}} by parts using the boundary conditions $G(r_c) = 0$
and $f_n(r_c) = 0$ to get
\begin{align}
f_n(r) = \,\, & f_c^{\prime} r_c^{-2(n+1)} \int_{r_c}^{r} u^{2(n+1)} du - \nonumber \\
& \frac{1}{2n+3} \left(r^{2n+3} G(r) - \int_{r_c}^{r} u^{n+2} g_n(u) du\right).
\label{eq:fsoln1}
\end{align}

If we now make use of \Eq{\ref{eq:fc_prime}} to eliminate the unknown $f_c^{\prime}$, and perform the first
integral, we obtain
\begin{align}
f_n(r) = \,\, & \frac{1}{2n+3}\left(r^{2n+3}-r_c^{2n+3}\right)G(\infty)-\nonumber \\
& \frac{1}{2n+3} \left(r^{2n+3} G(r) - \int_{r_c}^{r} u^{n+2} g_n(u) du\right).
\label{eq:fsoln1a}
\end{align}

Since, by definition $G(\infty) - G(r) = \int_r^{\infty} g_n u^{-(n+1)} du$, we can combine the two terms with factor
$r^{2n+3}$ and multiply both sides by $r^{-(n+1)}$ to get the following form for the full radial part of the solution
\begin{align}
f_n(r) r^{-(n+1)}  = \,\, & \frac{1}{2n+3} \left[ r^{n+2} \int_{r}^{\infty} u^{-(n+1)}
g_n(u) du + \right. \nonumber \\
                  &  r^{-(n+1)} \left(\int_{r_c}^{r} u^{n+2} g_n(u) du -
\right.     \nonumber \\
                  &   \left. \left.   r_c^{2n+3} \int_{r_c}^{\infty}
u^{-(n+1)} g_n(u) du \right) \right].
\label{eq:fsoln2}
\end{align}

We have written the solution in this form so that it reflects the full radial part of the solution given in
\Eq{\ref{eq:fullalpha}}. This radial part of the full solution agrees
with that given by \citet{caudal} and
\citet{lackner1970}. Their work shows that the integral multiplied by a
factor $r_c^{2n+3}$ comes about by assuming a boundary condition for
$f_n^\prime$ different from zero. However, Caudal points out that this extra
integral corresponds to surface currents at $r=r_c$ and makes negligible
contribution to the solution beyond a few planetary radii. In fact for his
final calculations he omits it and relies on a more detailed internal field
model. For the work described in this paper, we use a simple centred
dipole representation of Saturn's field, with equatorial field strength as given
in \Table{\ref{scales_table}}.


The final solution consists of the homogeneous part (assumed to be the dipole or other appropriate potential) added to the particular solution (non-zero source) 
whose radial and angular parts we have derived above. For completeness, we now give here the final solution for the magnetodisc potential
\begin{align}
\alpha(r,\mu)  = \,\, & \frac{1-\mu^2}{r} + \nonumber  \\
& (1-\mu^2)\sum_{n=0}^{\infty}\frac{\jacpoly{n}{\mu}}{2n+3}
\left[r^{n+2}\int_{r}^{\infty}g_n(u)u^{-(n+1)}du\right. \nonumber \\
& + r^{-(n+1)} \left(\int_{r_c}^{r} u^{n+2} g_n(u) du - \right. \nonumber \\
& \left.\left. r_c^{2n+3}\int_{r_c}^\infty u^{-(n+1)}g_n(u)du\right)\right].
\label{eq:formal_mdisc_soln}
\end{align}

This represents, in practice, a cumbersome calculation. 
The number of terms required in the polynomial series depends on how accurate a 
representation is needed for the source function (whether empirical or theoretical).
Source functions characterised by larger angular scales require fewer polynomials in the expansion. 
For the work described in this paper, we used polynomial expansion up to degree $n=30$.
The corresponding latitudinal resolution captured by the polynomial of this degree
is $\sim2.2^{\circ}$, corresponding to typical vertical resolutions at the equator
in the range \unit[0.2\mbox{--}1]{\RS}. To obtain final model outputs, we stopped
iteration when the maximum relative difference in the solution for the magnetic potential $\alpha$ 
between consecutive iterations became less than 0.5 percent.

\section{Scaling of physical quantities}
\label{scales_appendix}
\Table{\ref{scales_appendix_table}} presents a summary of the scaling values
for all dimensionless quantities.
\begin{table}
\centering
\caption{Scaling values between variables in physical units and
their dimensionless counterparts for both planets.}
\label{scales_appendix_table}
\begin{tabular}[b]{llrrr}
\hline 
Dimension & Definition & Units & Saturn & Jupiter \\
\hline
\multicolumn{5}{c}{\emph{Primary scales}}\\
Length & $a$ & \unit{km} & 60280 & 71492 \\
Magnetic Field & $\Beq$ & \unit{nT} &  21160 & 428000 \\
\hline
\multicolumn{5}{c}{\emph{Derived scales}}\\
Volume & $a^3$ & \unit{km^3} & $2{\cdot}10^{14}$ & $4{\cdot}10^{14}$ \\
Magnetic Flux & $\Beq a^2$ & \unit{GWb} & 77 & 2187 \\
Current density & $\Beq/(a\mu_0)$ & \unit{nA\,m^{-2}} & 280 & 4800 \\
Pressure & $\Beq^2/\mu_0$ & \unit{Pa} & 0.00036 & 0.146 \\
Energy density & $\Beq^2/\mu_0$ & \unit{J\,m^{-3}} & 0.00036  & 0.146  \\
Hot plasma index $K_h$ & $\Beq a/\mu_0$ & \unit{Pa\,m\,T^{-1}} &
$10^9$ & $2\cdot10^{10}$ \\
\hline
\end{tabular}  
\end{table}

\section{Derivation of the plasma scale length}
\label{sec:scale_length}
This appendix describes the derivation of the plasma scale length $\lsc$
defined in \Eq{\ref{eq:ldefn}}.
Let us first consider the force balance along a magnetic field line assuming
that the ions and the electrons are subject to an ambipolar electric potential
$\Phi_{\|}$.
In these conditions the force balance equation for the ions and the electrons
are written respectively 
\begin{align}
-\deriv{P_\|}{s}+nm_i\omega^2\rho\cos\varphi-ne\deriv{\Phi_\|}{s}&=0,
\label{eq:force_balance_ion}\\
-\deriv{P_\|}{s}+nm_e\omega^2\rho\cos\varphi+ne\deriv{\Phi_\|}{s}&=0.
\label{eq:force_balance_electron}
\end{align}
Here $s$ represents the curvilinear coordinate along a magnetic field line
(i.e.\ such that the potential $\alpha(s)$ is constant) and is oriented
toward the equator (i.e.\ such that positive force means force acting toward
the equator). The angle $\varphi$ denotes the angle between the magnetic
field line (in direction of increasing $s$) and the radial direction (with unit vector
$\vec{e}_\rho$).  
In these force balance equations we have implicitly assumed quasi-neutrality
of the plasma (i.e.\ $n_i=n_e=n$), and also assumed that ion and electron
pressure along the magnetic field are equal (i.e.\ 
$P_{\|_i}=P_{\|_e}=P_\|=n\kb T_\|$).  This assumption is consistent
with the definition for total pressure used by \citet{caudal} for a quasi-neutral
plasma with the specific ion charge number $Z=1$. However, it is worth noting
that this simplifying assumption would have to be relaxed for more realistic studies
of the observed
differences between ion and electron pressures. 
Subtracting \Eq{\ref{eq:force_balance_electron}} from 
\Eq{\ref{eq:force_balance_ion}} we obtain the following
expression for the ambipolar electric field $E_\|$, which is directed along the magnetic field 
\begin{align}
E_\|=-\deriv{\Phi_\|}{s}\approx-\frac{1}{2}\frac{m_i}{e}\omega^2\rho\cos\varphi.
\end{align}
We note that $E_\|$ is negative which means that the
ions tend to be `lifted' off the equator to a greater degree than they would be in the
absence of ambipolar effects.
Substituting back the expression for the ambipolar electric field $E_\|$
into the ion force balance equation \Eq{\ref{eq:force_balance_ion}},
we obtain the following differential equation
\begin{align}
\deriv{P_\|}{s} = -\frac{1}{2}
\frac{P_\|\rho\cos\varphi}{\kb T_\|/(m_i\omega^2)}.
\end{align}
Changing the curvilinear coordinate $s$ into the cylindrical radial distance
$\rho$ (i.e.\ $d\rho = ds\cos\varphi$ along the field line) we obtain the separable differential
equation
\begin{align}
\frac{dP_\|}{P_\|} = - \frac{\rho d\rho}{2\kb T_\|/(m_i\omega^2)}.
\end{align}
And finally expressing the radial distance $\rho$ in normalised units and
integrating from $\rho$ to the equatorial crossing $\rho_0$ (with pressure $P_{\|_0}$)
we obtain the following analytic expression for ion and electron pressure
along a magnetic field line as function of radial distance $\rho$
\begin{align}
P_\|(\rho) = P_{\|_0}\exp\left(-\frac{\rho^2-\rho^2_0}{2\lsc^2}\right),
\end{align}
where we recognise the plasma scale length $\lsc$ as given by
\Eq{\ref{eq:ldefn}}. Thus this definition of $\lsc$ in \citet{caudal}'s formalism
implicitly represents the above simplified expression for the ambipolar electric field in the plasma.
For our model, the charge state $Z=1$ and the value of $\lsc$ is a factor of $\sqrt{2}$ larger
than that which would be derived for ions in the absence of the ambipolar electric field.
For plasma ions with general charge number
$Z$, this factor is $\sqrt{Z+1}$.
As quasi-neutrality
is maintained, the electrons are also distributed with the same scale length $\lsc$ (since they have
negligible mass compared to the ions). A more thorough treatment of the polarisation electric field
in the magnetospheres of Jupiter and Saturn, along with its effects on plasma distributions, can be found
in \citet{maurice1997}. 

\label{lastpage}

\end{document}